\immediate\write18{makeindex \jobname.nlo -s nomencl.ist -o \jobname.nls}

\documentclass[3p]{elsarticle}

\UseRawInputEncoding
\usepackage{tabularx}
\usepackage{amsmath}
\usepackage{amssymb}
\usepackage{graphicx}
\usepackage{subfigure}
\usepackage{enumerate}
\usepackage{float}
\usepackage[percent]{overpic}
\usepackage{sidecap}
\usepackage{tipa}
\usepackage{epstopdf}
\usepackage{booktabs}
\usepackage{natbib}
\usepackage{hyperref}
\usepackage{lineno}
\usepackage{appendix}
\usepackage{arydshln}
\usepackage{mathtools}
\usepackage{slashbox}
\usepackage{diagbox}

\usepackage{scalerel,stackengine,amsmath}
\newcommand\equalhat{\mathrel{\stackon[1.5pt]{=}{\stretchto{%
    \scalerel*[\widthof{=}]{\wedge}{\rule{1ex}{3ex}}}{0.5ex}}}}
\newcommand\norm[1]{\left\lVert#1\right\rVert}

\usepackage{tikz}

\modulolinenumbers[5]

\DeclareMathOperator{\SALI}{SALI}
\DeclareMathOperator{\clicks}{Clicks}

\usepackage{framed} 
\usepackage[nomentbl,stdsubgroups,nocfg]{nomencl} 

\makenomenclature

\makeatletter
\def\ps@pprintTitle{%
  \let\@oddhead\@empty
  \let\@evenhead\@empty
  \let\@oddfoot\@empty
  \let\@evenfoot\@oddfoot
}
\makeatother

\usepackage{numcompress}\biboptions{authoryear}

\begin{document}

\begin{frontmatter}

\title{Nonlinear dynamics and onset of chaos in a physical model of a damper pressure relief valve}



\author[mainaddress]{Lukas Schickhofer\corref{correspondingauthor}}
\address[mainaddress]{KTM Research \& Development, Mattighofen, Austria}
\cortext[correspondingauthor]{Corresponding author:}
\ead{lukas.schickhofer13@alumni.imperial.ac.uk}

\author[secondaryaddress]{Chris G. Antonopoulos}
\address[secondaryaddress]{Department of Mathematical Sciences, University of Essex, Wivenhoe Park, United Kingdom}

\begin{abstract}
Hydraulic valves, for the purpose of targeted pressure relief and damping, are a ubiquitous part of modern suspension systems.
This paper deals with the analytical computation of fixed points of the dynamical system of a single-stage shock absorber valve, which can be applied for the direct computation of its system variables at equilibrium without brute-force numerical integration.
The obtained analytical expressions are given for the original dimensional version of the model derived from continuity and motion equations for a realistic valve setup. Furthermore, a large part of the study is devoted to a systematic sensitivity analysis of the model towards crucial parameter changes. Numerical investigation of a potential loss of stability and following nonlinear oscillations is performed in multi-dimensional parameter spaces, which reveals sustained valve vibrations at increased valve mass and applied pretension force. The dynamical behaviour is analysed by phase space orbits, as well as Fourier-transformed valve displacement data to identify dominant frequencies.
Chaotic indicators, such as Lyapunov exponents and the Smaller Alignment Index (SALI), are utilized to understand the nature of the oscillatory motion and to distinguish between order and chaos.
\end{abstract}

\begin{keyword}
Non-smooth dynamical system\sep Impact oscillator\sep Stability analysis\sep Parametric study
\MSC[2010] 00-01\sep  99-00
\end{keyword}

\end{frontmatter}

\section{\textbf{Introduction}}
\label{Sec:1}

It is long known that shock absorbers and automotive dampers, which contain various forms of mechanical valves subjected to fluid pressure, show complex and sometimes unexpected behaviour, such as instabilities, or sustained vibrations and noise during operation. This is mainly due to the dominant physical nonlinearities, which are also present in the underlying equations of motion.
These nonlinear terms come in several forms, such as coupled pressure-displacement relationships, impact laws, or variable spring stiffness, which collectively render the dynamical system non-smooth and therefore difficult to analyse with usual methods of nonlinear dynamics \citep{kunze2001non,di2008bifurcations}.

Although early works, especially by \citet{lang1977study}, \citet{reybrouck1994non}, and \citet{talbott2002experimentally}, introduced potent models for the prediction of shock absorber variables, these approaches were either algebraic and linearized in nature and therefore not suited for studies of its nonlinear dynamics, or they applied a large set of semi-empirical fitting parameters that did not allow investigation of the actual physical effects and co-dependencies of real model properties. \citet{duym1997evaluation} also gives a critical evaluation of early damper models, stating some of their weaknesses and shortcomings, such as the problems in capturing hysteresis and comparable nonlinear effects.
Consequently, models based on nonlinear differential equations and actual material properties and physical parameters are required for the analysis of realistic cause-effect relationships between model properties and behaviour, as well as its dynamics.
More recently, \citet{farjoud2012nonlinear} devised a set of nonlinear differential equations for the modelling of hydraulic damper valves and successfully predicted their opening characteristics at different shim stack and orifice settings. 
\citet{boggs2010efficient} and \citet{sikora2018modeling} applied similar analytic methods to model the suspension damping behaviour in real-world vehicle applications. Furthermore, \citet{liao2021fluid} derived at suitable two-degree-of-freedom model for fluid-structure coupling within a relief valve for the parameter optimization of underwater applications.

As a result of ongoing research and development in pressure relief valve dynamics, particularly in the automotive industry, it became necessary to understand instability mechanisms and the loss of dynamic stability of hydraulic valves in operation.
In an early numerical investigation of instabilities in hydraulic valve systems, \citet{hayashi1997chaos} could identify the Feigenbaum route to chaos by power spectra, bifurcation diagrams and Lyapunov exponents for a simplified poppet control valve.
\citet{benaziz2012nonlinear} applied a first-order differential equation system to predict the dynamics of spring valves and studied the global stability of hydraulic valves under realistic parameter changes. Thus, they could map out the occurrence of instabilities under various flow rates. Additionally, they provided a nonlinear model for the analysis of valve vibration and the resulting structure-borne noise \citep{benaziz2015shock}.
Using various approaches like the brute-force numerical integration of long time series of trajectories with implicit schemes, as well as bifurcation analyses of non-impacting periodic solutions, \citet{hHos2012grazing} and \citet{bazso2014bifurcation} were able to identify initial instabilities and various bifurcation types, such as fold, torus, and grazing bifurcations, in a non-dimensional model of a pressure relief valve. 
Recently, an experimental study by \citet{ma2019experimental} confirmed these instability mechanisms of pressure relief valves and specified the occurrence of chatter instabilities by general stability maps for a wide range of volumetric flow rates and geometric properties such as inlet pipe length.
The valve types in these studies are similar to the one studied here, but lack the energy dissipation of a viscous fluid surrounding the valve.
Furthermore, to date no systematic investigation of the effect of parameter variations on the global dynamics of suspension valves has been attempted. One reason for this was the non-existence of a validated closed system of differential equations suitable for the application of numerical integration and fixed-point analysis.

Recently, such a physical model was introduced and successfully applied to analyse inertia and dynamic stability of a typical automotive suspension valve system \citep{schickhofer2022fluid,schickhofer2022universal}.
In this work, its derivative in the form of a coupled, dimensional nonlinear system of equations for a shock absorber pressure relief valve is studied with respect to its equilibrium states and its sensitivity to parameter changes. The chosen approach is therefore a mixture of analytical methods for the derivation of formulae for direct computation, as well as numerical methods for the integration of the first-order system.
Moreover, to understand whether the system undergoes changes in the nature of its dynamics from order to chaos, or vice versa, chaotic indicators such as Lyapunov exponents \citep{benettin1980lyapunov1,benettin1980lyapunov2} and the Smaller Alignment Index (SALI) \citep{skokos2003does,skokos2004detecting} are applied. Although the studied system is generally dissipative, as explained in detail for variations of the damping parameter in Sec. \ref{Sec:3C-5}, it could well display transient chaotic motion with subsequent loss of function.
It is therefore important, both from a fundamental, as well as an applied perspective on pressure relief valves and shock absorber systems, to understand and accurately predict such transitions and sensitivities.

Various model parameters are known to have a significant impact on the shock absorber valve function and dynamic behaviour. However, little is known on their precise impact on equilibrium states, transition to chaos, and instabilities. This study attempts to answer three important questions: 
\begin{itemize}
\item Is it possible to find analytic expressions for the equilibrium states of the main shock absorber variables, such as valve displacement, velocity, and pressure?
\item How are the nonlinear valve dynamics affected by realistic parameter changes?
\item Can we distinguish between order and chaos in the motion of the investigated dissipative dynamical system?
\end{itemize}
In the following, we will address these these questions through analytical and numerical methods and we will return to them in Sec. \ref{Sec:4} for a detailed discussion.

\section{\textbf{Theory}}
\label{Sec:2}

\begin{figure}[htbp]
\begin{centering}
\begin{overpic}[width=0.80\textwidth]{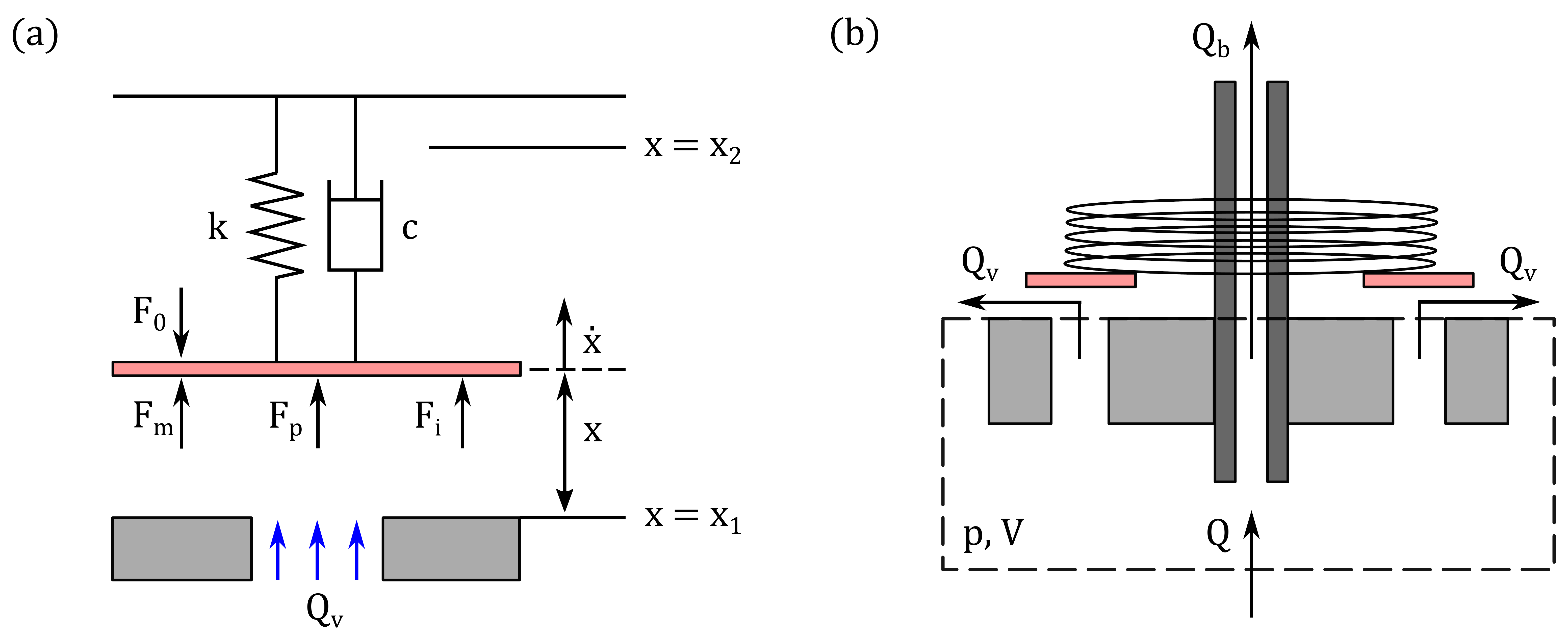}
\end{overpic}
\par
\end{centering}
\caption{Diagram of pressure force $F_{p}$, momentum force $F_{m}$, impact force $F_{i}$, and pretension $F_{0}$ acting on the shock absorber valve (a), and sketch of the control volume $V$ involving the total volumetric flow rate $Q$, valve flow rate $Q_{v}$, and bleed flow rate $Q_{b}$ (b) used for the fully integrated form of the continuity equation (\ref{Eq:2A-2}). 
A detailed description and definition of forces and flow rates is given in Sec. \ref{Sec:2A}.
The boundaries at $x=x_{1}$ and $x=x_{2}$ of the domain and valve motion are indicated.}
\label{Fig:2-1} 
\end{figure}

The dynamical system to be studied is the three-dimensional model of a valve for the purpose of pressure relief in a hydraulic shock absorber.
A shock absorber acts as part of the suspension system of various types of vehicles, such as cars, motorcycles, or trains, to absorb wheel displacements as a result of ground irregularities or impact. It contains a viscous fluid, typically mineral oil, which is compressed by the motion of a rod in a piston. Its valve is further displaced by the acting pressure. The pressure drop across the valve initiates flow rates through a piston with orifices of various cross sections, most notably the flow rate $Q_{b}$ through a constant bleed orifice and the flow rate $Q_{v}$ through a variable valve opening. This entire process leads to an efficient relief of pressure by various flow losses through the piston and valve gap flow. For a detailed description of the shock absorber concept, the reader is referred to \citet{dixon2008shock} and \citet{schickhofer2022universal}.
Figure \ref{Fig:2-1} shows the forces acting on the single valve and the volumetric flow rates across the considered control volume.
The underlying system of equations have been previously validated and successfully implemented for shock absorber development \citep{schickhofer2022fluid}.
Below, the dimensional version of the equation system is analysed with respect to its critical points.
\\

\subsection{\textbf{Dimensional system}}
\label{Sec:2A}

Essentially by applying Newton's second law on the valve shown in Fig. \ref{Fig:2-1}(a), and the mass continuity equation on the control volume depicted in Fig. \ref{Fig:2-1}(b), one arrives at the following equations: 
\begin{align}
m\ddot{x}+c\dot{x}+kx &= F_{p}\left(p\right)+F_{m}\left(x,p\right)+F_{i}\left(x\right)-F_{0},  \label{Eq:2A-1}  \\
\dot{p} &= \frac{1}{\beta V}\left[Q-Q_{v}\left(x,p\right)-Q_{b}\left(p\right)\right].  \label{Eq:2A-2} 
\end{align}
Equation (\ref{Eq:2A-1})(a) models the valve as a driven oscillator of mass $m$, damping $c$, and spring constant $k$, which is subjected to the forces given below.
The mineral oil used in the damper is assumed to be compressible with effective compressibility $\beta$, and the integrated continuity equation across the volume $V$ takes into account all flow rates introduced in Fig. \ref{Fig:2-1}(b).

\paragraph{Pressure force}
The dominant force is given by the static pressure difference $\Delta p=p-p_{0}$ across the valve between the variable upstream pressure $p$ and the downstream 
pressure $p_{0}$ of the shock absorber piston. It depends crucially on the imposed volumetric flow rate further described below and causes a pressure force given by
\begin{equation}
F_{p}\left(p\right)=A_{p}\Delta p=A_{p}\left(p-p_{0}\right) \equalhat A_{p}p.
\label{Eq:2A-3} 
\end{equation}
Here, $A_{p}$ is the total area on the valve on which the fluid pressure acts. In many shock absorbers this area consists of the sum of several port sections.
If the reference pressure is considered zero ($p_{0}=0$), or if $p$ is assumed as gauge pressure, the expression in Eq. (\ref{Eq:2A-3}) simplifies to $A_{p}p$.

\paragraph{Momentum force}
Due to the redirection of the fluid jet by the valve surface, a momentum change is taken into account by the force term
\begin{equation}
F_{m}\left(x,p\right)=\frac{C_{f}\rho}{A_{p}}\left(\underbrace{\alpha\pi d_{v}x}_{A_{v}\left(x\right)}C_{d,v}\sqrt{\frac{2p}{\rho}}\right)^2,
\label{Eq:2A-4} 
\end{equation}
which is dependent on the valve opening $x$ and pressure $p$, but also on the empirically determined momentum coefficient $C_{f}$, which controls the relative amount of transferred momentum. 
\citet{lang1977study} established this parameter through experiments at $C_{f}\approx0.3$, but it can vary significantly between shock absorber geometries due to differences in piston geometry design and the resulting flow paths, which directly affect the impact angle of the jet on the valve surface and thus the effective area \citep{hHos2014dynamic}.
The valve opening section $A_{v}\left(x\right)=\alpha P x = \alpha\pi d_{v}x$ is a direct function of the product of perimeter and valve displacement, which gives a generated surface for the flow. The flow proportionality coefficient $\alpha$ is used to adjust the actual flow area depending on the geometry.
An alternative definition of the momentum force, $F_{m}=\dot{m}u_{0}$, can be given as a function of mass flow rate $\dot{m}$ and inlet mean flow velocity $u_{0}$.

\paragraph{Impact force}
The effect of impact at the lower and upper bounds $x_{1}$ and $x_{2}$ is modelled by a force term
\begin{equation}
F_{i}\left(x\right)=\begin{cases}
-k_{i}\left(x-x_{1}\right), & \quad x\leq x_{1},\\
0, & \quad x_{1}<x<x_{2},\\
-k_{i}\left(x-x_{2}\right), & \quad x\geq x_{2},
\end{cases}
\label{Eq:2A-5} 
\end{equation}
which models the repulsion using a stiff spring constant $k_{i}=\SI{1e10}{\newton\per\meter}$ that is considerably larger than the system stiffness ($k_{i} \gg k$).

\paragraph{Pretension force}
The constant force $F_{0}$ can be controlled via pretension of the spring acting on the valve.
Its purpose is the targeted delay of opening during initial pressure increase.

\paragraph{Volumetric flow rates}
The total volumetric flow rate $Q$ is introduced to the system by displacement due to road irregularities.
Furthermore, the definitions of the valve flow rate
\begin{equation}
Q_{v}\left(x,p\right)=C_{d,v}\underbrace{\alpha \pi d_{v}x}_{A_{v}\left(x\right)}\sqrt{\frac{2\Delta p}{\rho}},
\label{Eq:2A-6} 
\end{equation}
and of the flow rate through the constant bleed orifice 
\begin{equation}
Q_{b}\left(p\right)=C_{d,b}A_{b}\sqrt{\frac{2\Delta p}{\rho}},
\label{Eq:2A-7} 
\end{equation}
follow directly from the Bernoulli equation, which is considered to accurately describe the flow across the piston under the premise that the discharge coefficients $C_{d,v}$ and $C_{d,b}$ are accurately chosen to reflect the transition from laminar to turbulent flow, as suggested by \citet{segel1981mechanics}.
As introduced through Fig. \ref{Fig:2-1}, the bleed orifice is a constriction of constant cross section in the centre of the shock absorber piston, while the valve lift leads to a variable opening. The effect of both paths is a cumulative flow rate and pressure drop across the shock absorber.

Using the above definitions, Eq. (\ref{Eq:2A-1})-(\ref{Eq:2A-2}) can be rewritten as a three-dimensional, first-order system
\begin{align}
\dot{y}_{1} &= y_{2},  \label{Eq:2A-8a}  \\
\dot{y}_{2} &= \frac{1}{m} \left[-cy_{2}-ky_{1}+F_{p}\left(y_{3}\right)+F_{m}\left(y_{1},y_{3}\right)+F_{i}\left(y_{1}\right)-F_{0}\right],  \label{Eq:2A-8b}  \\
\dot{y}_{3} &= \frac{1}{\beta V} \left[Q-Q_{v}\left(y_{1},y_{3}\right)-Q_{b}\left(y_{3}\right)\right],  \label{Eq:2A-8c} 
\end{align}
where $\vec{y}=\left(y_{1},y_{2},y_{3}\right)=\left(x,\dot{x},p\right)$.

The system (\ref{Eq:2A-8a})-(\ref{Eq:2A-8c}) can subsequently be used for numerical integration within a large parameter space. By varying the flow rate $Q$ and certain valve properties, the dynamics and transitions from stable to unstable states are investigated in Sec. \ref{Sec:3B} and Sec. \ref{Sec:3C}.
Numerical integration is carried out by schemes capable of dealing with stiff equations and non-smooth systems, such as ODE15s or ODE23s in \textit{MATLAB}. For the data displayed in multi-dimensional parameter spaces, such as the plots depicting valve displacement as a function of flow rate and model parameters in Sec. \ref{Sec:3C}, task farming on a computing cluster is applied, which allows for the parallel integration of many single-core cases for sufficiently long runtimes.

\subsection{\textbf{Chaotic indicators}}
\label{Sec:2B}

\subsubsection{\text{Lyapunov exponents}}
\label{Sec:2B-1}

Lyapunov characteristic exponents (LCE) are a well-known tool to gauge the rate of separation from an orbit point in $\mathbb{R}^m$ along its $m$ orthogonal directions.
In three dimensions, this leads to a set of three Lyapunov exponents, 
\begin{equation}
L_{k}=\lim_{t\to\infty}{\left(\frac{1}{t}\ln{\left[\frac{\vec{w}_k\left(t\right)}{\vec{w}_k\left(0\right)}\right]}\right)}=\lim_{t\to\infty}{\left(\frac{1}{t}\ln{\left[\vec{r}_k\left(t\right)\right]}\right)},
\label{Eq:2A-8d} 
\end{equation}
indicating the shrinking or stretching of orthogonal axes $\vec{r}_k\left(t\right)$ that span the state space with $k=1,\ldots,m$ at time $t$. If the largest Lyapunov exponent settles to a non-zero, positive value, this gives an indication of chaotic motion, since infinitesimally close initial conditions result in orbits that diverge exponentially fast in time \citep{yorke1996chaos}.

\subsubsection{\text{Smaller Alignment Index (SALI)}}
\label{Sec:2B-2}

An alternative to Lyapunov exponents for the detection of chaos is the Smaller Alignment Index (SALI) \citep{skokos2003does,skokos2004detecting}, which is defined as
\begin{equation}
\SALI\left(t\right)=\min\left\{\norm{ \frac{\vec{w}_1\left(t\right)}{\norm{\vec{w}_1\left(t\right)}}+\frac{\vec{w}_2\left(t\right)}{\norm{\vec{w}_2\left(t\right)}} }, \norm{ \frac{\vec{w}_1\left(t\right)}{\norm{\vec{w}_1\left(t\right)}}-\frac{\vec{w}_2\left(t\right)}{\norm{\vec{w}_2\left(t\right)}} }\right\}.
\label{Eq:2A-8e} 
\end{equation}
It uses the fact that any two randomly chosen initial deviation vectors $\vec{w}_1\left(0\right)$, $\vec{w}_2\left(0\right)$ will eventually become aligned with the most unstable direction with the angle between them tending to zero. Since one is only interested in the direction of the deviation vectors, they are only normalized at each timestep and not orthogonalized, which is a crucial difference to the computation of Lyapunov exponents and saves computing time. In the case of chaotic orbits the normalized vectors align and point to the same (or exactly opposite) direction. Thus the norm of their sum (\emph{antiparallel alignment index}) or difference (\emph{parallel alignment index}) tends to zero.
It follows from Eq. (\ref{Eq:2A-8e}) that generally $\SALI\left(t\right)\in\left[0,\sqrt{2}\right]$, with $\lim_{t\to\infty}{\SALI\left(t\right)}=0$ being a clear indication of chaos.
SALI provides information on the instantaneous changes of the dynamics and is therefore an important additional quantity for chaotic indication. While Lyapunov exponents show the long-term persistent trend of a dynamical system either towards order or chaos, in particular after convergence and considerable integration time, SALI is able to identify also transient chaos and is especially relevant for the analysis of short-lived oscillations of realistic, dissipative systems, such as the one studied in this paper.

\subsection{\textbf{Validation}}
\label{Sec:2C}

\begin{figure}[htbp]
\begin{centering}
\begin{overpic}[width=0.50\textwidth]{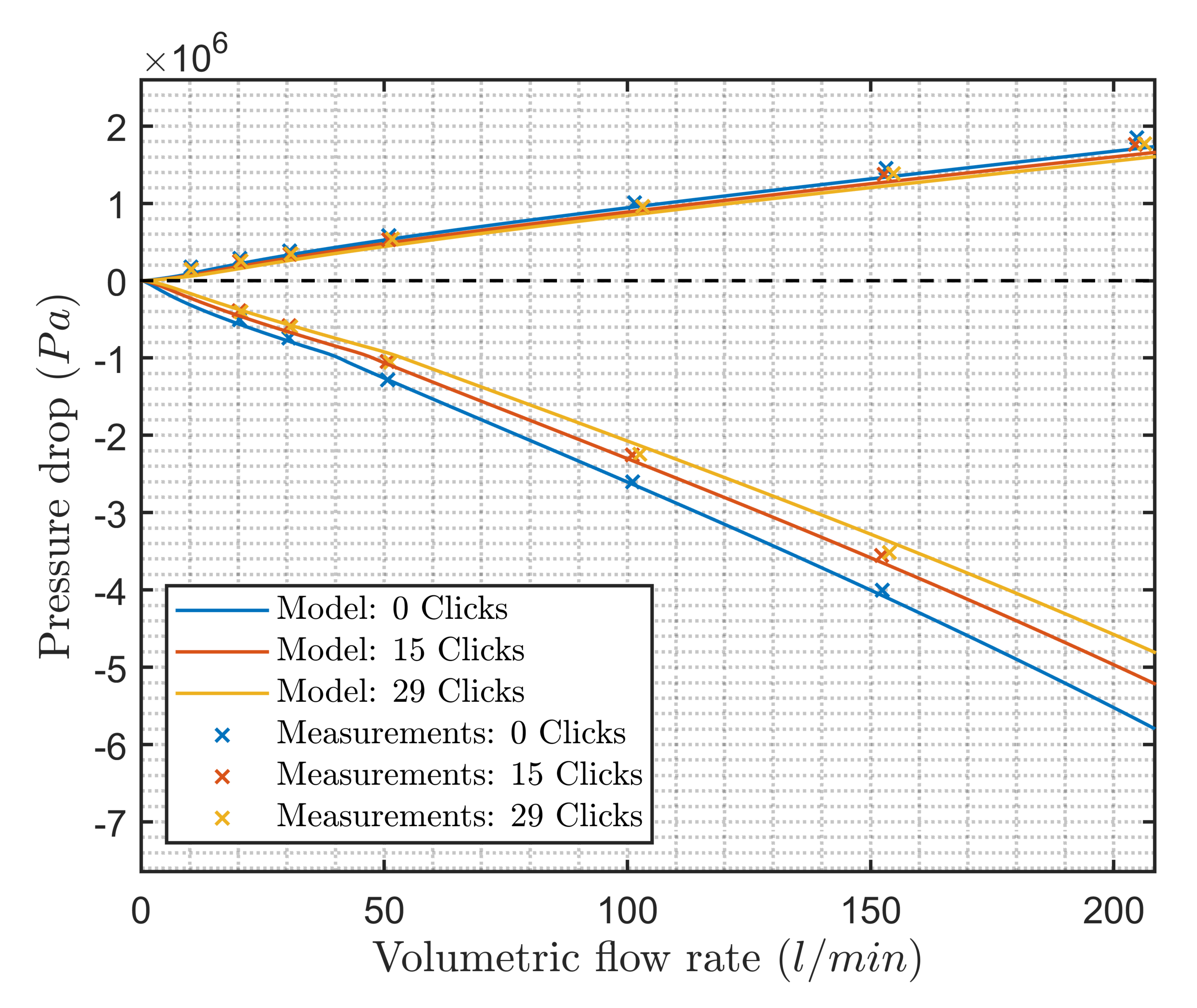}
\end{overpic}
\par
\end{centering}
\caption{Numerically computed pressure loss of a realistic damper including a pressure-relief valve modelled by the system (\ref{Eq:2A-8a})-(\ref{Eq:2A-8c}) compared against test bench measurements.}
\label{Fig:2-2} 
\end{figure}

Extensive validation of the model (\ref{Eq:2A-8a})-(\ref{Eq:2A-8c}) has been carried out both against fluid-structure interaction (FSI) simulation data in \citet{schickhofer2022fluid} as well as against experimental test bench data of damping curves in \citet{schickhofer2022universal}. 
A typical approach in suspension engineering has been carried out in these measurements: Using a time-resolved dynamometer with controlled sinusoidal mass flow input of ground excitation $w_{g}=2\pi f_{g}$,
\begin{equation}
\dot{m}\left(t\right) = \dot{m}_{max}\left(t\right)\sin\left(\omega_{g}t\right), \nonumber
\end{equation}
the pressure drop across the hydraulic shock absorber valve has been measured and compared to the pressure output of the model. Here, a reference frequency of ground excitation of $f_{g}=\SI{8}{\hertz}$ has been chosen. Since the pressure reduction across the valve system is directly coupled to the other model variables, such as valve opening displacement, this constitutes a sufficient setup for valve model validation. Figure \ref{Fig:2-2} shows a comparison of numerically obtained pressure loss from the system (\ref{Eq:2A-8a})-(\ref{Eq:2A-8c}) against test bench data of a realistic shock absorber including a hydraulic pressure relief valve of $0$, $15$, and \SI{29}{\clicks}, where \SI{}{\clicks} refers to a common configuration unit for the adjustment of the bleed orifice area (${A}_{b}=\SI{0}{\square\milli\meter}$  at \SI{0}{\clicks}, ${A}_{b}=\SI{1.4}{\square\milli\meter}$  at \SI{15}{\clicks}, ${A}_{b}=\SI{2.7}{\square\milli\meter}$  at \SI{29}{\clicks}).
The other baseline model parameters of the check valve are given by the values from \ref{App:A}. It can be seen that the agreement between the model's prediction and the test bench measurements is satisfactory and lies consistently below $\SI{4}{\percent}$ for the compression stroke of positive pressure drop values and the rebound stroke of negative values in Fig. \ref{Fig:2-2}.

\section{\textbf{Results}}
\label{Sec:3}

In the following, both direct analytical approaches, as well as numerical integration of the first-order system (\ref{Eq:2A-8a})-(\ref{Eq:2A-8c}) are performed to investigate the characteristics of the valve dynamics. While Sec. \ref{Sec:3A} offers a mathematical formulation for the immediate computation of equilibrium states in the valve motion, Sec. \ref{Sec:3B}-\ref{Sec:3B} explore the physical consequences of crucial parameter changes of the system.
The three-dimensional dynamical system, as introduced by Eq. (\ref{Eq:2A-8a})-(\ref{Eq:2A-8c}) is solved with the values for its model parameters as presented in \ref{App:A}. Wherever specific parameters are varied, this is noted in text and figures. 

\subsection{\textbf{Fixed points}}
\label{Sec:3A}

The investigation of fixed points of a dynamical system is valuable for a number of reasons:
First of all, they give the long-term state to which the system converges after all initial transients have disappeared and constitute important stable solutions for a given set of initial and boundary conditions. Additionally, if analytical expressions can be found for these critical points, no numerical integration is necessary for their computation, which is an important mathematical property.

In order to find the fixed points of the system (\ref{Eq:2A-8a})-(\ref{Eq:2A-8c}), its gradients must vanish such that
\begin{equation}
\dot{\vec{y}}_{e} = \left. \frac{\partial\vec{y}}{\partial t}\right|_{\vec{y}_{e}} = \vec{f}\left(\vec{y}_{e}\right) = \vec{0},
\label{Eq:2A-9} 
\end{equation}
which gives
\begin{align}
y_{2} &= 0,  \label{Eq:2A-10a}  \\
-ky_{1}+F_{p}\left(y_{3}\right)+F_{m}\left(y_{1},y_{3}\right)+F_{i}\left(y_{1}\right)-F_{0} &= 0,  \label{Eq:2A-10b}  \\
Q-Q_{v}\left(y_{1},y_{3}\right)-Q_{b}\left(y_{3}\right) &= 0.  \label{Eq:2A-10c} 
\end{align}
Equation (\ref{Eq:2A-10a}) simply enforces zero velocity of the valve at equilibrium, while Eq. (\ref{Eq:2A-10c}) recovers essentially the conservation of volume across the control domain.
Thus, there will be no compressibility effects at fixed points of the system.
As a result, Eq. (\ref{Eq:2A-10b}) is analysed as the main equation of state involving valve opening $y_{1}$ and acting pressure $y_{3}$:
\begin{align}
y_{1} &= \frac{1}{k}\left(F_{p}\left(y_{3}\right)+F_{m}\left(y_{1},y_{3}\right)+F_{i}\left(y_{1}\right)-F_{0}\right)  \nonumber \\
          &= \frac{1}{k}\left[A_{p}y_{3}+\frac{C_{f}\rho}{A_{p}}\left(\alpha\pi d_{v}C_{d,v}y_{1}\sqrt{\frac{2y_{3}}{\rho}}\right)^2-\underbrace{k_{i}\left(y_{1}-x_{1/2}\right)}_{\forall y_{1} \leq x_1 \land y_{1} \geq x_{2}}-F_{0}\right].  \label{Eq:2A-11} 
\end{align}
This is a quadratic equation that can be rewritten to give an expression for $y_{1}$.
By rearranging and defining constant coefficients, we obtain
\begin{align}
\underbrace{\frac{2C_{f}}{kA_{p}}\left(\alpha\pi d_{v}C_{d,v}\right)^2}_{a}y_{1}^{2}y_{3}-\left(1+\frac{k_{i}}{k}\right)y_{1}+\underbrace{\frac{A_{p}}{k}}_{b}y_{3}+\underbrace{\frac{k_{i}}{k}}_{c}\underbrace{x_{1/2}}_{d}-\underbrace{\frac{F_{0}}{k}}_{e} &= 0,  \nonumber
\end{align}
which can be rewritten as
\begin{equation}
ay_{3}y_{1}^2-\left(1+c\right)y_{1}+by_{3}+cd-e = 0.   
\label{Eq:2A-12} 
\end{equation}
Equation (\ref{Eq:2A-10c}) can be further applied to get a relationship between displacement $y_{1}$ and pressure $y_{3}$:
\begin{align}
y_{3} &= \frac{\rho}{2} \left(\frac{Q}{C_{d,v}\alpha\pi d_{v}y_{1}+C_{d,b}A_{b}}\right)^{2}  \nonumber \\
          &= \underbrace{\frac{\rho Q^{2}}{2}}_{h} \frac{1}{\left(\underbrace{C_{d,v}\alpha\pi d_{v}}_{f}y_{1}+\underbrace{C_{d,b}A_{b}}_{g}\right)^{2}},  \nonumber
\end{align}      
which results in    
\begin{equation}         
y_{3} = \frac{h}{\left(fy_{1}+g\right)^{2}}.  
\label{Eq:2A-13} 
\end{equation}
By using Eq. (\ref{Eq:2A-13}) for Eq. (\ref{Eq:2A-12}), one ends up with an equation of third order in $y_{1}$:
\begin{align}
ay_{1}^2\frac{h}{\left(fy_{1}+g\right)^{2}} - \left(1+c\right)y_{1} + b\frac{h}{\left(fy_{1}+g\right)^{2}} + cd - e &= 0,  \nonumber \\
ahy_{1}^2 - \left(1+c\right)\left(fy_{1}+g\right)^{2}y_{1} + \left(cd - e\right)\left(fy_{1}+g\right)^{2} + bh &= 0.  \label{Eq:2A-14}
\end{align}
In order to alleviate the analytical treatment, three cases are distinguished that reflect typical realistic conditions of the shock absorber:

\paragraph{Case 1: $F_{i} = 0$, $F_{0} = 0$}
For motion of the valve with vanishing impact force and pretension, this case gives a reasonable approximation. It leads to the following expression:
\begin{align}
ahy_{1}^{2} - \left(fy_{1}+g\right)^{2}y_{1} + bh = 0,  \nonumber  \\
f^{2}y_{1}^{3} + \left(2fg-ah\right)y_{1}^{2} + g^{2}y_{1} - bh = 0,  \nonumber  \\
y_{1}^{3} + \underbrace{\frac{2fg-ah}{f^{2}}}_{A}y_{1}^{2} + \underbrace{\frac{g^{2}}{f^{2}}}_{B}y_{1} - \underbrace{\frac{bh}{f^{2}}}_{C} = 0.   \label{Eq:2A-15} 
\end{align}
Thus, the resulting relationship can be written as a third-order equation in $y_{1}$ of the type $y_{1}^{3}+Ay_{1}^{2}+By_{1}-C=0$, which can be solved by
\begin{align}
y_{1} \approx & 0.26457 \sqrt[3]{-2A^{3}+5.1962\sqrt{-4A^{3}C-A^{2}B^{2}+18ABC+4B^{3}+27C^{2}}+9AB+27C}  \nonumber  \\
& - \frac{0.41997\left(3B-A^{2}\right)}{\sqrt[3]{-2A^{3}+5.1962\sqrt{-4A^{3}C-A^{2}B^{2}+18ABC+4B^{3}+27C^{2}}+9AB+27C}}  \nonumber  \\
& - 0.33333A.
\label{Eq:2A-15a} 
\end{align}

\paragraph{Case 2: $F_{i} = 0$, $F_{0} \neq 0$}
If pretension force is applied and regular smooth motion without impact is considered, we get
\begin{align}
ahy_{1}^{2} - \left(fy_{1}+g\right)^{2}y_{1} - e\left(fy_{1}+g\right)^{2} + bh = 0,  \nonumber  \\
f^{2}y_{1}^{3} + \left(2fg-ah+ef\right)y_{1}^{2} + \left(g^{2}+2efg\right)y_{1} - bh +eg^{2} = 0,  \nonumber  \\
y_{1}^{3} + \underbrace{\frac{2fg-ah+ef}{f^{2}}}_{A}y_{1}^{2} + \underbrace{\frac{g^{2}+2efg}{f^{2}}}_{B}y_{1} - \underbrace{\frac{bh-eg^{2}}{f^{2}}}_{C} = 0,   \label{Eq:2A-16} 
\end{align}
which can be solved in a similar fashion as above:
\begin{align}
y_{1} \approx & 0.26457 \sqrt[3]{-2A^{3}+5.1962\sqrt{-4A^{3}C-A^{2}B^{2}+18ABC+4B^{3}+27C^{2}}+9AB+27C}  \nonumber  \\
& - \frac{0.41997\left(3B-A^{2}\right)}{\sqrt[3]{-2A^{3}+5.1962\sqrt{-4A^{3}C-A^{2}B^{2}+18ABC+4B^{3}+27C^{2}}+9AB+27C}}  \nonumber  \\
& - 0.33333A.
\label{Eq:2A-16a} 
\end{align}

\paragraph{Case 3: $F_{i} \neq 0$, $F_{0} \neq 0$}
For the case of acting impact and pretension forces, which involves all terms of Eq. (\ref{Eq:2A-14}), we can attempt a comparison of the order of present forces. Since the stiffness of the impact force term is several orders of magnitude larger than the system's stiffness ($k_{i}/k \approx 10^{6}$), we can approximate Eq. (\ref{Eq:2A-14}) by
\begin{equation}
-c\left(y_{1}+d\right)=0,
\label{Eq:2A-17} 
\end{equation}
which leads to the fixed points 
\begin{equation}
y_{1} = d = x_{1/2},
\label{Eq:2A-18} 
\end{equation}
at the lower and upper bound of the impact oscillator (cf. Fig. \ref{Fig:2-1}). These are pseudo-equilibria of the non-smooth dynamical system, which are defined exclusively by the discontinuities of the domain.

\begin{figure}[htbp]
\begin{centering}
\begin{overpic}[width=1.00\textwidth]{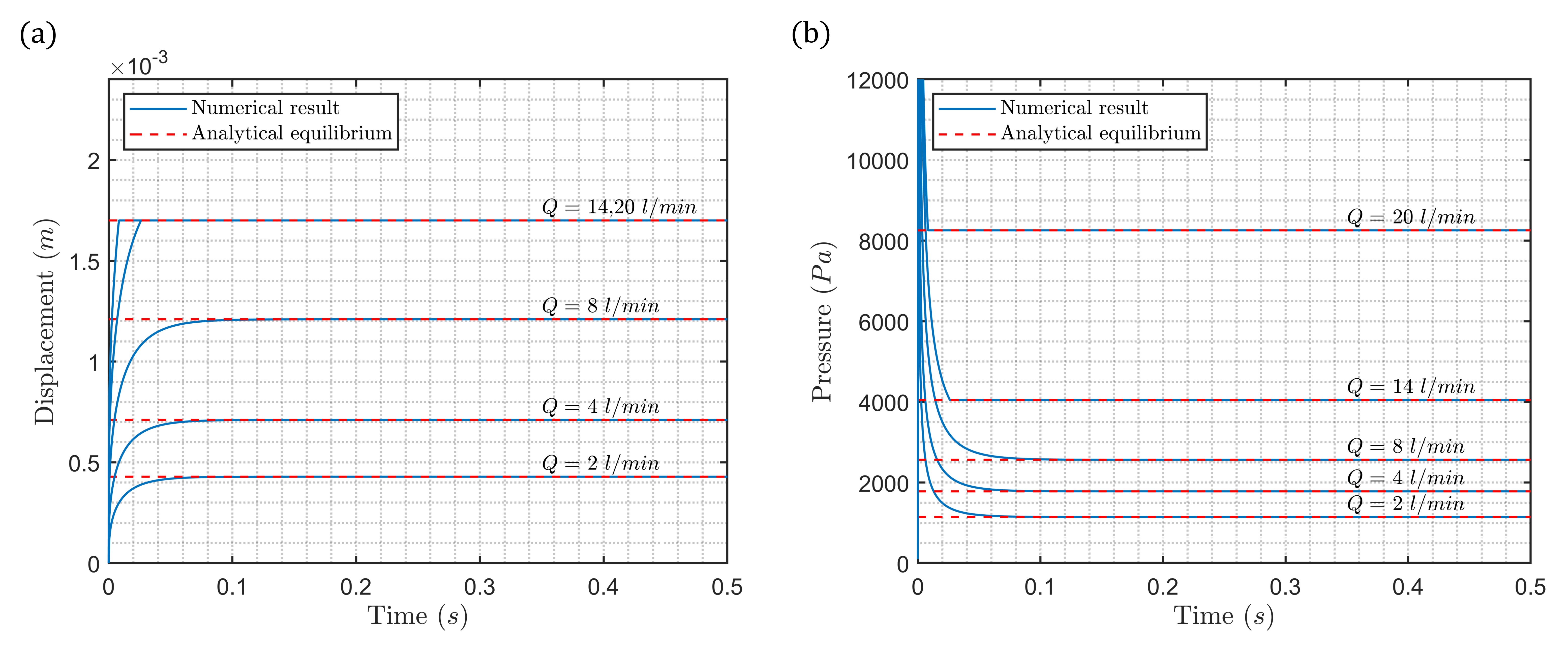}
\end{overpic}
\par
\end{centering}
\caption{Analytically computed equilibrium solutions for displacement and pressure compared to the results from numerical integration of the system (\ref{Eq:2A-8a})-(\ref{Eq:2A-8c}) for typical volumetric flow rates of $[2,4,8,14,20]\,\SI{}{\liter\per\min}$.}
\label{Fig:3A-1} 
\end{figure}

With the analytically obtained expressions for equilibrium displacements of Eq. (\ref{Eq:2A-15a}) and Eq. (\ref{Eq:2A-18}), together with the definition of Eq. (\ref{Eq:2A-13}) for the related pressure, one has the possibility of computing those valve parameters without numerically integrating system (\ref{Eq:2A-8a})-(\ref{Eq:2A-8c}).
Figure \ref{Fig:3A-1} shows a comparison of analytical and numerical solutions for relevant flow rates. After initial transients have subsided, the time-dependent numerical result settles to an equilibrium and the two solutions are identical.

\subsection{\textbf{Valve opening characteristics}}
\label{Sec:3B}

An important aspect of valve dynamics is the opening behaviour during increase of the flow rate. 
In this section, we investigate the possibility of popping-off instabilities or valve chattering during opening, as well as the occurrence of oscillations during impact at the boundaries of the non-smooth system (\ref{Eq:2A-8a})-(\ref{Eq:2A-8c}). The impact force (\ref{Eq:2A-5}) is acting at $x_{1}=\SI{0}{\milli\meter}$ and $x_{2}=\SI{1.7}{\milli\meter}$ (cf. \ref{App:A}) in this setup, which are valve seat values for shock absorber applications.

\begin{figure}[htbp]
\begin{centering}
\begin{overpic}[width=1.00\textwidth]{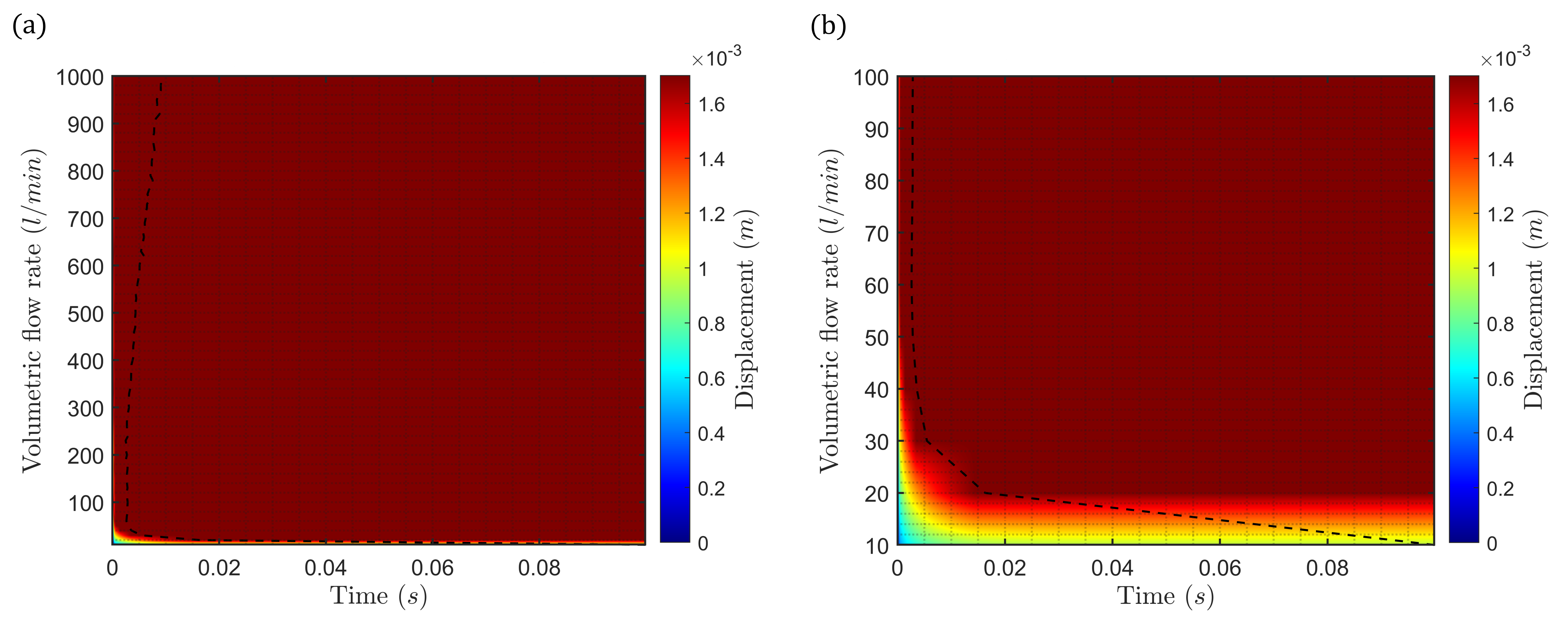}
\end{overpic}
\par
\end{centering}
\caption{Dynamics of valve opening for a range of $0$--$\SI{1000}{\liter\per\min}$ (a) and $0$--$\SI{100}{\liter\per\min}$ (b). The time needed for reaching the opening amplitude at $\SI{1.7}{\milli\meter}$ is shown together with a black dashed line indicating the point at which the solution has settled to equilibrium and all gradients of the first-order system (\ref{Eq:2A-8a})-(\ref{Eq:2A-8c}) have vanished.}
\label{Fig:3B-1} 
\end{figure}

Depending on the flow velocity magnitudes, the valve might show delays in settling to an equilibrium state. This effect is demonstrated in Fig. \ref{Fig:3B-1} and occurs due to initial vibrations and further violent chattering during closing, especially at intermediate to high flow rates. Initially, there are persistent instabilities at low volume flow due to the fact that the forces acting on the valve are of similar order of magnitude (cf. Eq. (\ref{Eq:2A-1})). After recovery to a constant delay of equilibrium above approximately $\SI{50}{\liter\per\min}$, it increases linearly with the applied volume flow at higher flow rates at above $\SI{200}{\liter\per\min}$.
Thus, there is a clear indication of an ideal range of operation of the shock absorber valve with respect to the fluid flow it can handle. Predictable and largely constant opening characteristics are obtained for volumetric flow rates of $60$--$\SI{190}{\liter\per\min}$.

\subsection{\textbf{Parametric studies}}
\label{Sec:3C}

The impact of various parameter changes is studied below, where crucial quantities such as the bleed orifice area, mass, or pretension are varied (cf. Sec. \ref{Sec:2A}).
All parameter ranges are chosen based on a realistic check valve model as it would occur in a hydraulic shock absorber and using the baseline properties from \ref{App:A}.

\subsubsection{\text{Bleed orifice area}}
\label{Sec:3C-1}

\begin{figure}[htbp]
\begin{centering}
\begin{overpic}[width=1.00\textwidth]{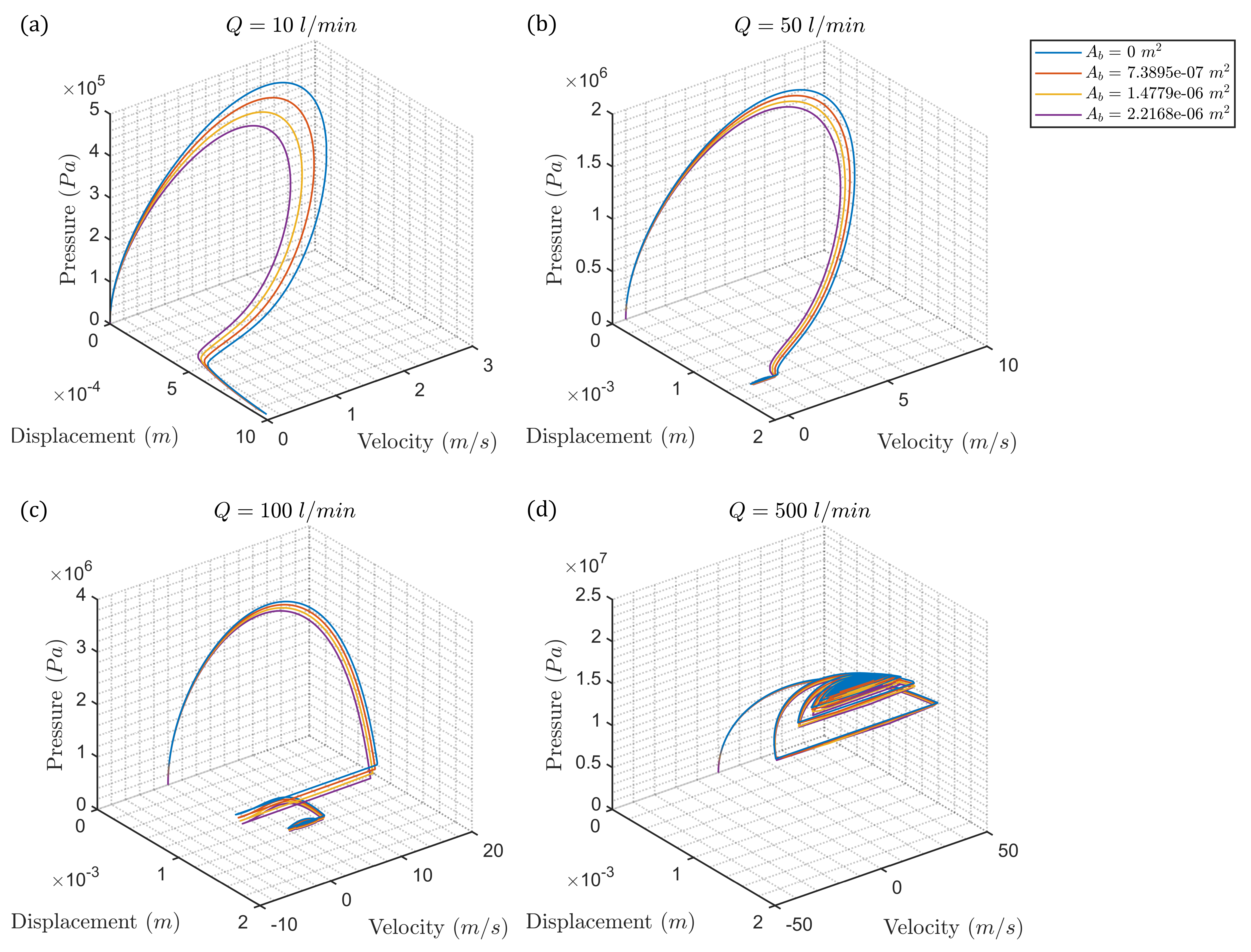}
\end{overpic}
\par
\end{centering}
\caption{Orbits in phase space of displacement, velocity, and pressure for specific values of the bleed orifice area and increasing flow rates of $Q=\SI{10}{\liter\per\minute}$ (a), $Q=\SI{50}{\liter\per\minute}$ (b), $Q=\SI{100}{\liter\per\minute}$ (c), and $Q=\SI{500}{\liter\per\minute}$ (d).}
\label{Fig:3C-1} 
\end{figure}

The cross-sectional area of the bleed orifice regulates the flow through the constant orifice of the valve, as detailed in Eq. (\ref{Eq:2A-7}). It thereby generates a pressure loss due to flow constriction and subsequent expansion, which occurs in parallel to the flow losses through the spring-loaded valve.
Fig. \ref{Fig:3C-1} reflects the fact that changes in the bleed orifice area have a linear effect on the valve dynamics. The orbits change only slightly in magnitude, but the underlying motion pattern stays the same. Therefore, all dynamic motion captured in Fig. \ref{Fig:3C-1} can be considered stable.
 
\subsubsection{\text{Mass}}
\label{Sec:3C-2}

\begin{figure}[htbp]
\begin{centering}
\begin{overpic}[width=1.00\textwidth]{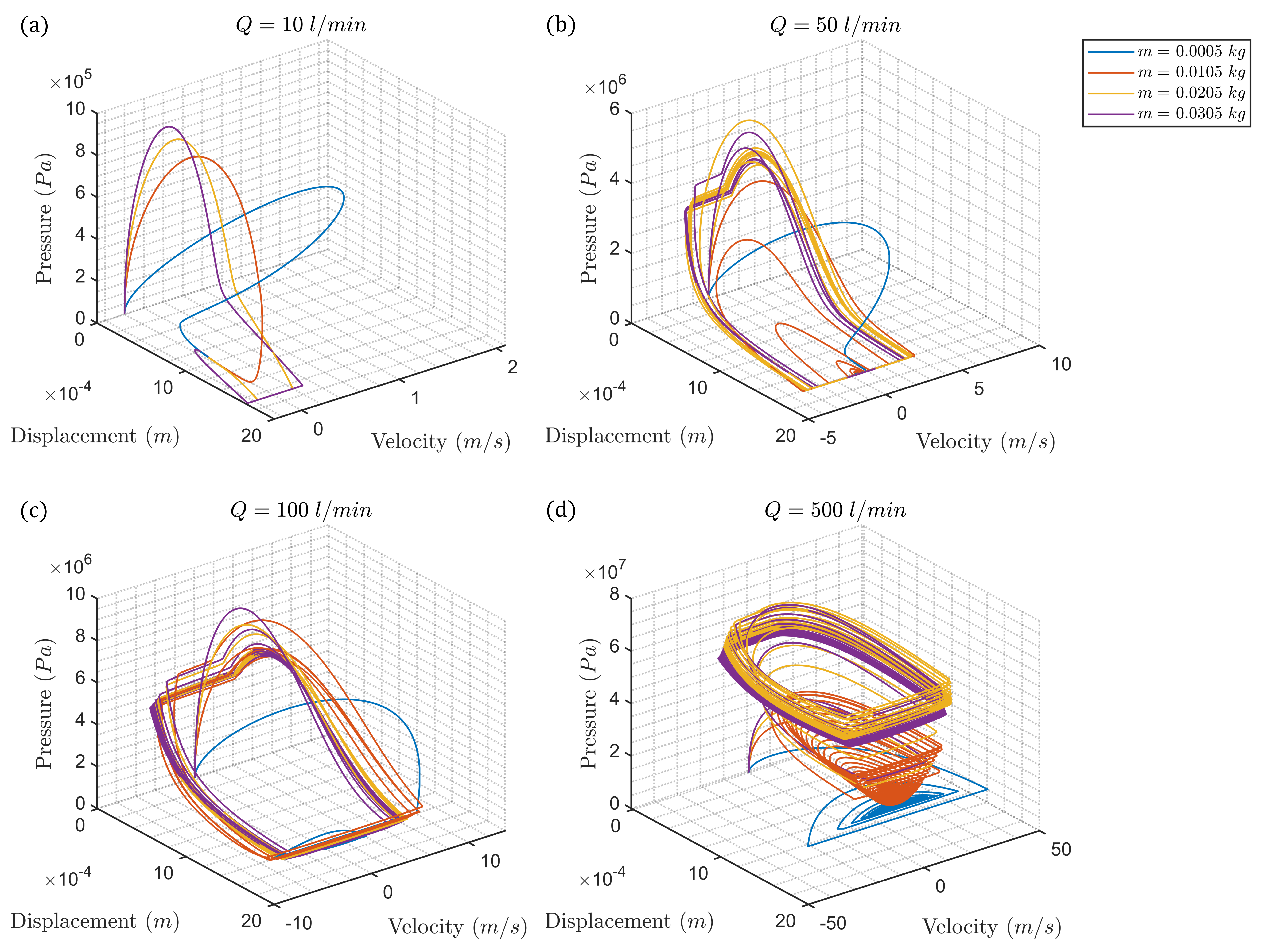}
\end{overpic}
\par
\end{centering}
\caption{Phase space with trajectories for various valve masses and increasing flow rates of $Q=\SI{10}{\liter\per\minute}$ (a), $Q=\SI{50}{\liter\per\minute}$ (b), $Q=\SI{100}{\liter\per\minute}$ (c), and $Q=\SI{500}{\liter\per\minute}$ (d).}
\label{Fig:3C-2} 
\end{figure}

\begin{figure}[htbp]
\begin{centering}
\begin{overpic}[width=1.00\textwidth]{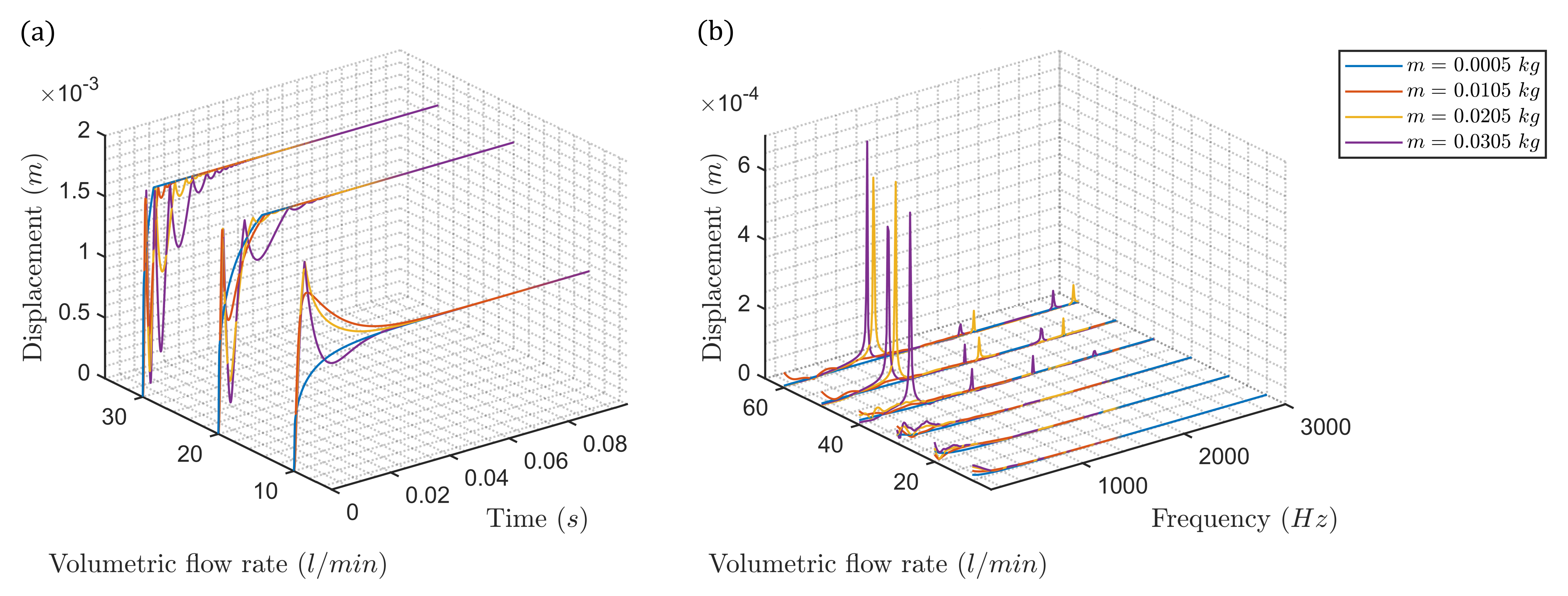}
\end{overpic}
\par
\end{centering}
\caption{Displacement time series (a) and Fourier transformation (b) for realistic volumetric flow rates and the chosen mass values.}
\label{Fig:3C-3} 
\end{figure}

\begin{figure}[htbp]
\begin{centering}
\begin{overpic}[width=1.00\textwidth]{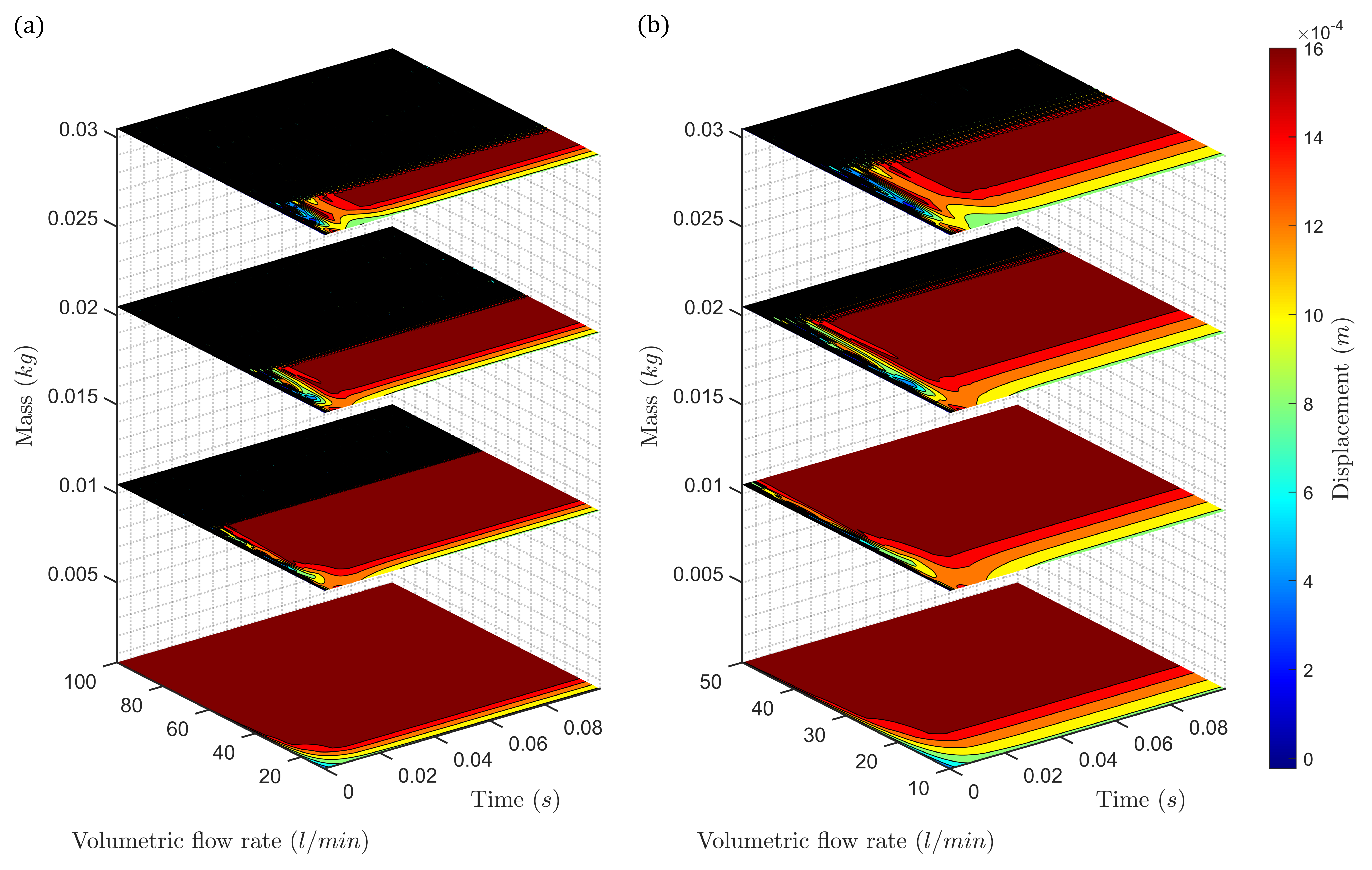}
\end{overpic}
\par
\end{centering}
\caption{Parameter space of valve displacement as a function of mass, time, and flow rate in a range of $0$--$\SI{100}{\liter\per\min}$ (a) and $0$--$\SI{50}{\liter\per\min}$ (b).}
\label{Fig:3C-4} 
\end{figure}

A crucial parameter for the dynamic behaviour of a pressure relief valve is its mass. It can be seen from Fig. \ref{Fig:3C-2} that an increase of mass has a potentially destabilizing effect due to larger inertia. This leads to violent \emph{chattering}, a boundary effect of non-smooth dynamical systems and impact oscillators, which entails a series of impacts that cause vibrations of the valve in real applications \citep{mora2013non}. Table \ref{Tab:1} indicates the fundamental stability properties of the valve at different mass and flow rates.
This nonlinear phenomenon intensifies at higher flow rates and causes sustained loops, or even surfaces, along which the trajectories oscillate (cf. Fig. \ref{Fig:3C-2}(b)-(d)).
The tendency of the valve to show stable oscillations thereby clearly increases at higher mass, as further demonstrated by the Fourier transformation of the displacement time series data shown in Fig. \ref{Fig:3C-3}. Additionally, the sections through the parameter space in Fig. \ref{Fig:3C-4} reveal ranges of volumetric flow rates of sustained oscillations visualized in black that extend to lower flow rates within the sections of higher masses. Furthermore, initial instabilities during valve opening are bleeding into lower flow rates at intermediate mass, as indicated by the isolated patches of higher displacement in Fig. \ref{Fig:3C-4}(b).

\begin{table}[htbp]
\caption{Stability map summarizing the results of Fig. \ref{Fig:3C-2} for various mass values and flow rates, where the symbol $\times$ means \emph{stable} and $\circ$ means \emph{unstable}.}
\centering
\begin{tabular}{|l||*{4}{c|}} \hline
\diagbox{Flow rate $Q$}{Mass $m$} & \SI{0.0005}{\kilo\gram} & \SI{0.0105}{\kilo\gram} & \SI{0.0205}{\kilo\gram} & \SI{0.0305}{\kilo\gram} \\ \hline
\hline \centering \SI{10}{\liter\per\minute} & $\times$ & $\times$ & $\times$ & $\times$ \\
\hline \centering \SI{50}{\liter\per\minute} & $\times$ & $\circ$ & $\circ$ & $\circ$ \\
\hline \centering \SI{100}{\liter\per\minute} & $\times$ & $\circ$ & $\circ$ & $\circ$ \\
\hline \centering \SI{500}{\liter\per\minute} & $\times$ & $\circ$ & $\circ$ & $\circ$ \\ \hline
\end{tabular}
\label{Tab:1}
\end{table}

\subsubsection{\text{Pretension}}
\label{Sec:3C-3}

\begin{figure}[htbp]
\begin{centering}
\begin{overpic}[width=1.00\textwidth]{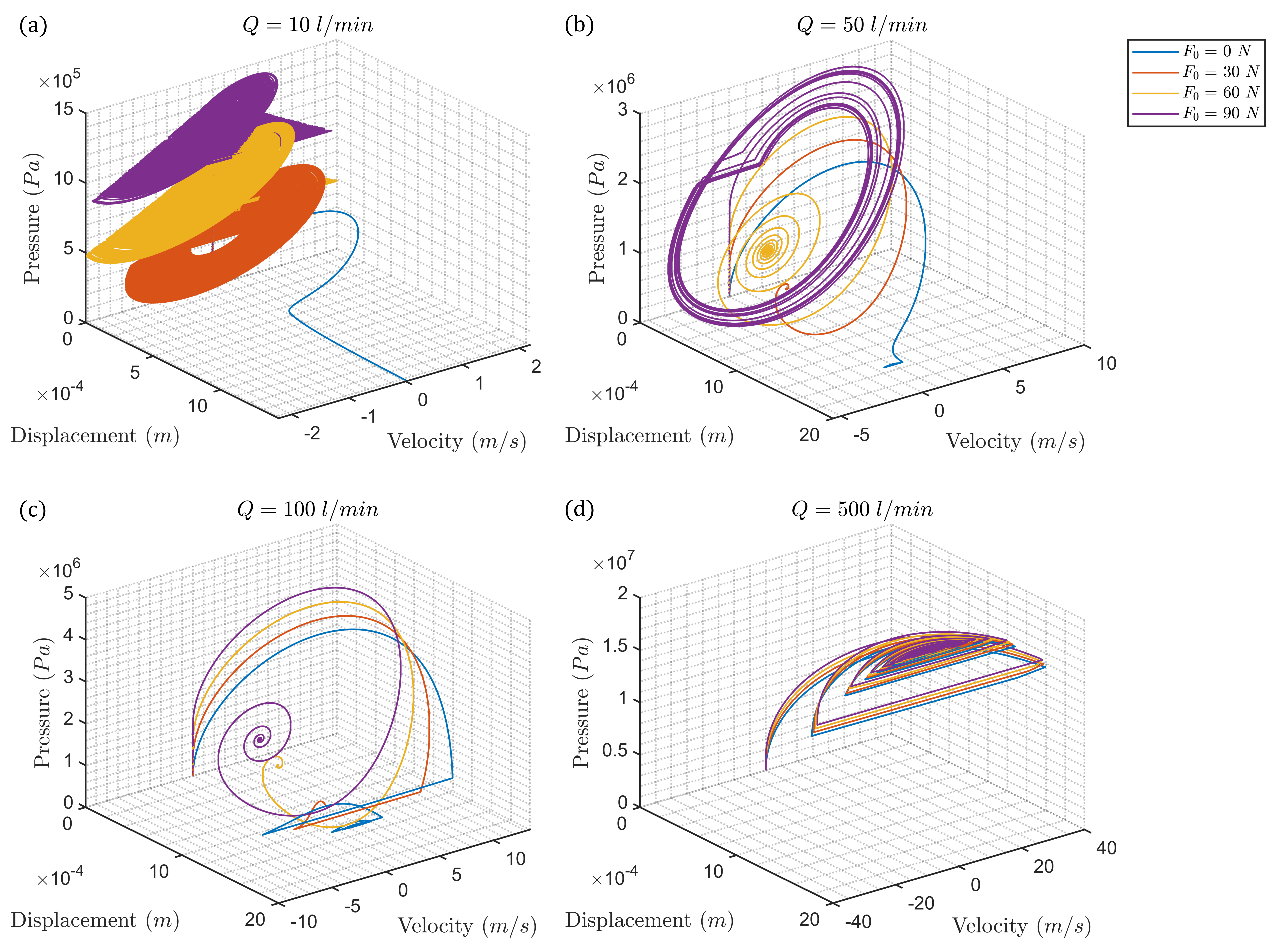}
\end{overpic}
\par
\end{centering}
\caption{Orbits in phase space of displacement, velocity, and pressure for specific values of pretension and increasing flow rates of $Q=\SI{10}{\liter\per\minute}$ (a), $Q=\SI{50}{\liter\per\minute}$ (b), $Q=\SI{100}{\liter\per\minute}$ (c), and $Q=\SI{500}{\liter\per\minute}$ (d).}
\label{Fig:3C-5} 
\end{figure}

\begin{figure}[htbp]
\begin{centering}
\begin{overpic}[width=1.00\textwidth]{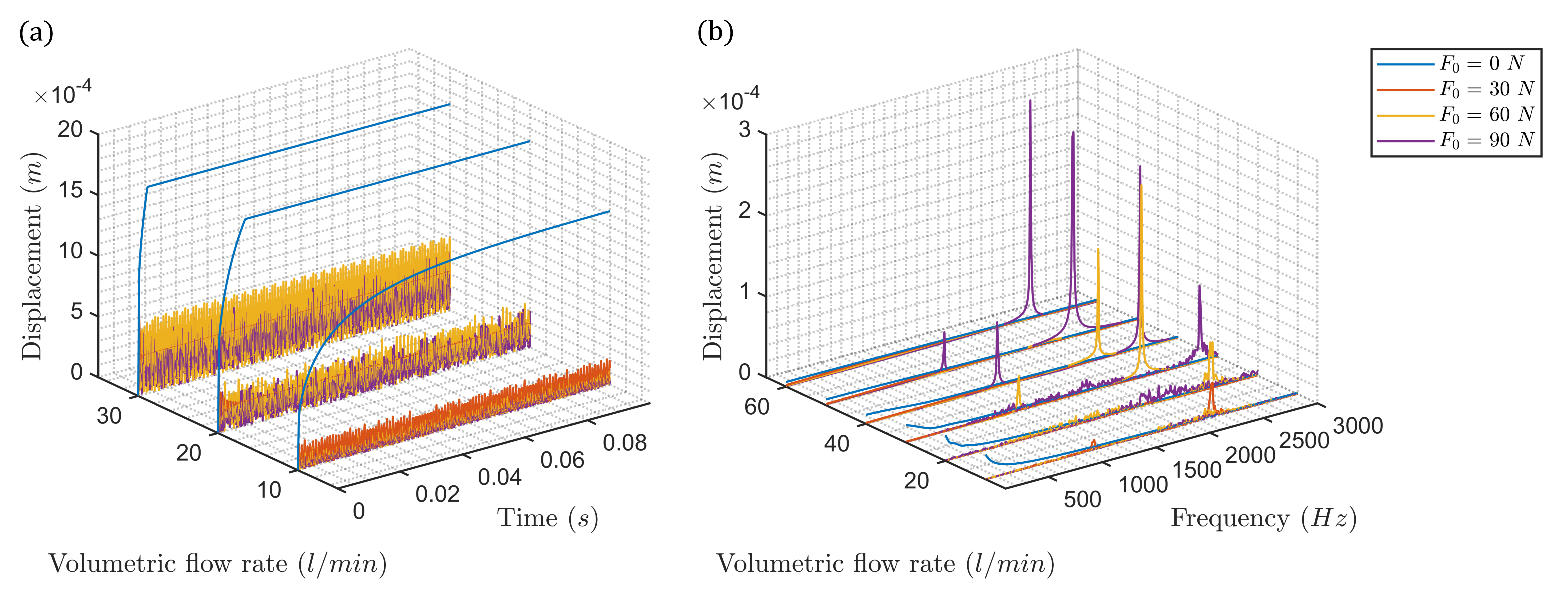}
\end{overpic}
\par
\end{centering}
\caption{Displacement in the time (a) and frequency domain (b) for a range of volumetric flow rates.}
\label{Fig:3C-6} 
\end{figure}

\begin{figure}[htbp]
\begin{centering}
\begin{overpic}[width=1.00\textwidth]{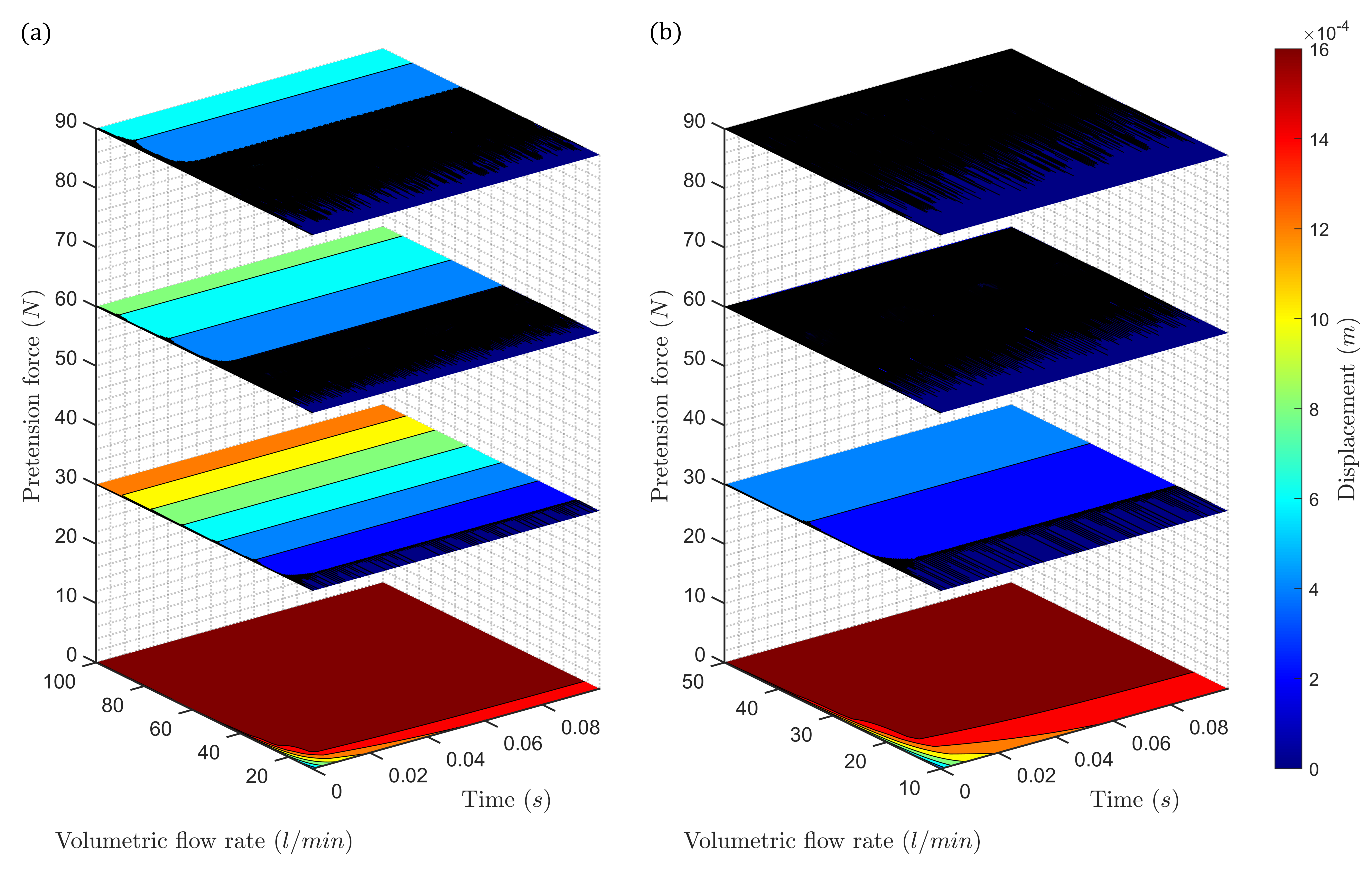}
\end{overpic}
\par
\end{centering}
\caption{Parameter space with displacement as a function of pretension, time, and volumetric flow rate in the range of $0$--$\SI{100}{\liter\per\min}$ (a) and $0$--$\SI{50}{\liter\per\min}$ (b).}
\label{Fig:3C-7} 
\end{figure}

The pretension force applied to a shock absorber valve delays opening and causes an overall higher pressure drop.
However, there is a problematic side effect of using a loaded spring for pretension: Particularly at lower volumetric flow rate and during the initial phase of opening there is a tendency to trigger instability and oscillations, which is visible in Fig. \ref{Fig:3C-5} and Tab. \ref{Tab:2}.
In Fig. \ref{Fig:3C-5}(a) the orbits in phase space exist on geometric structures similar in appearance to a chaotic attractor. Also Fig. \ref{Fig:3C-6} gives indications to aperiodic motion: While strong oscillatory behaviour can be recognized for all considered pretension force values, the motion undergoes a transition from relatively unordered vibrations without distinct frequencies at lower flow rates to regular periodic oscillations at frequencies of $\SI{2100}{}$--$\SI{2900}{\hertz}$. The transient character of this unwanted valve response becomes obvious by investigating the parameter spaces of Fig. \ref{Fig:3C-7}, where the black areas again indicate strong vibrations that are dominant when the terms for opposing forces in Eq. (\ref{Eq:2A-1}), such as the pressure force and momentum force on the one hand, and the pretension force on the other hand, become similar in magnitude. As soon as the volumetric flow rate is larger than a threshold value for the considered pretension, valve oscillations subside. This effect is also underlined by Tab. \ref{Tab:2}, which shows stable motion at intermediate to high flow rates above \SI{100}{\liter\per\minute} for all pretension values.

\begin{table}[htbp]
\caption{Stability map for the considered pretension values and flow rates of Fig. \ref{Fig:3C-5}, where $\times$ indicates a \emph{stable} setup and $\circ$ an \emph{unstable} one.}
\centering
\begin{tabular}{|l||*{4}{c|}} \hline
\diagbox{Flow rate $Q$}{Pretension $F_{0}$} & \SI{0}{\newton} & \SI{30}{\newton} & \SI{60}{\newton} & \SI{90}{\newton} \\ \hline
\hline \centering \SI{10}{\liter\per\minute} & $\times$ & $\circ$ & $\circ$ & $\circ$ \\
\hline \centering \SI{50}{\liter\per\minute} & $\times$ & $\circ$ & $\circ$ & $\circ$ \\
\hline \centering \SI{100}{\liter\per\minute} & $\times$ & $\times$ & $\circ$ & $\circ$ \\
\hline \centering \SI{500}{\liter\per\minute} & $\times$ & $\times$ & $\times$ & $\times$ \\ \hline
\end{tabular}
\label{Tab:2}
\end{table}

\begin{figure}[htbp]
\begin{centering}
\begin{overpic}[width=1.00\textwidth]{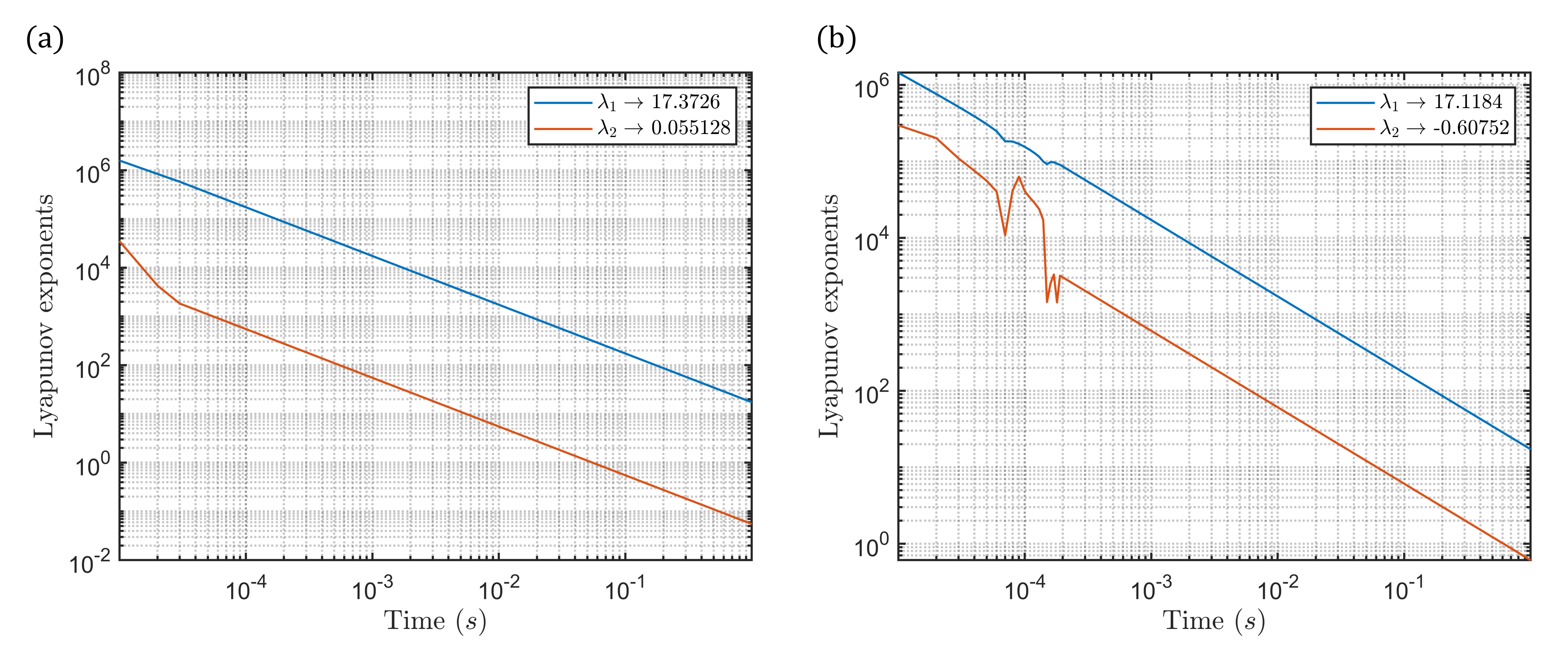}
\end{overpic}
\par
\end{centering}
\caption{Lyapunov exponents of the system at a pretension of $F_{0}=\SI{30}{\newton}$ (a) and $F_{0}=\SI{60}{\newton}$ (b) for $t \to \SI{1}{\second}$ in double-logarithmic plot.}
\label{Fig:3C-7a} 
\end{figure}

\begin{figure}[htbp]
\begin{centering}
\begin{overpic}[width=1.00\textwidth]{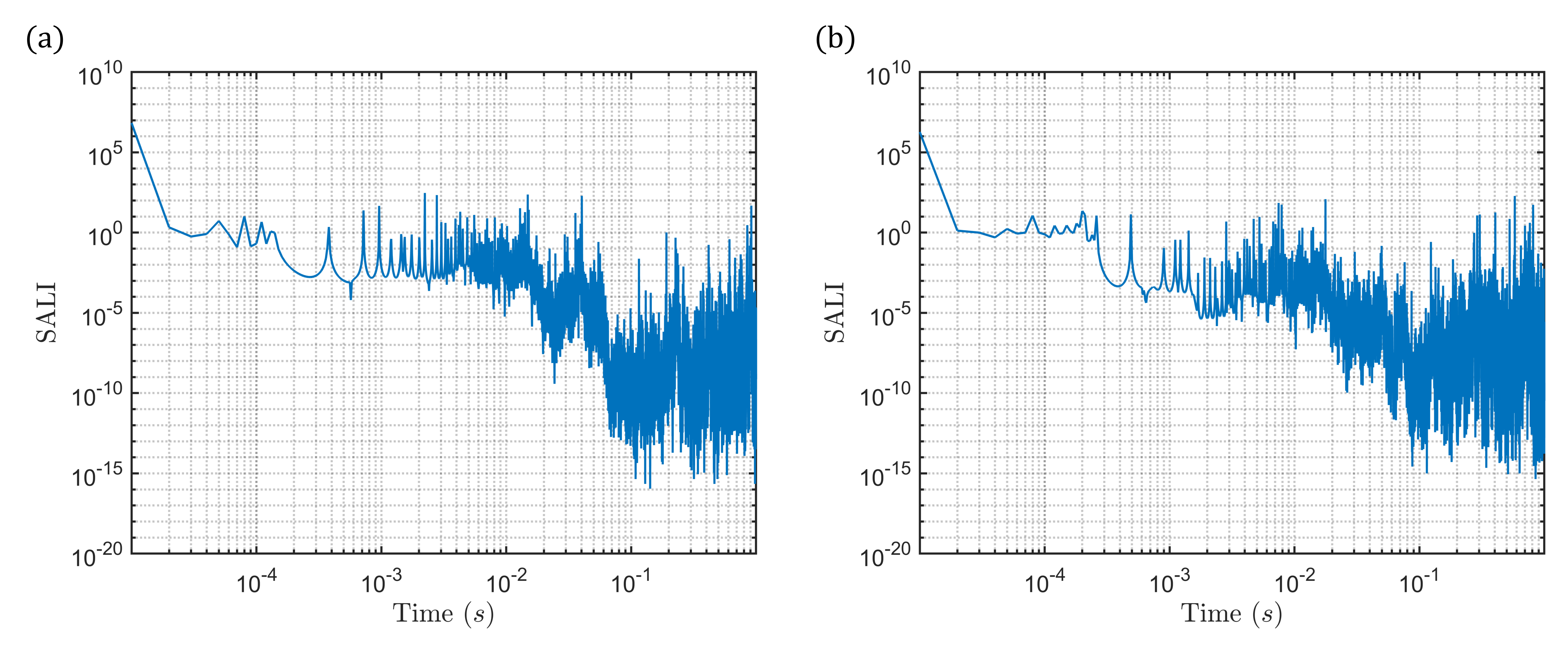}
\end{overpic}
\par
\end{centering}
\caption{Time evolution of SALI at a pretension of $F_{0}=\SI{30}{\newton}$ (a) and $F_{0}=\SI{60}{\newton}$ (b) for $t \to \SI{1}{\second}$ in logarithmic plot.}
\label{Fig:3C-7aa} 
\end{figure}

In order to further investigate the nature of the oscillatory motion described above, chaotic indicators are applied to the cases with pretension leading to orbits resembling chaotic trajectories. As visible in Fig. \ref{Fig:3C-5}, these are in particular pretensions of $F_{0}=\SI{30}{\newton}$ and $F_{0}=\SI{60}{\newton}$.
Figure \ref{Fig:3C-7a} shows the evolution of the Lyapunov exponents in time. It becomes clear from the double-logarithmic plot that it tends to zero. Nevertheless, intermediate ranges around $t=10^{-4}$ show a highly unsteady behaviour of the Lyapunov exponents, which are possibly related to transient chaos.
Moreover, the time series of SALI in Fig. \ref{Fig:3C-7aa} shows values that never fully settle below $\SALI\left(t\right) \leq 10^{-12}$. This confirms the notion of the consistently decreasing Lyapunov exponents from above and gives convincing reason to believe that the observed orbits of Fig. \ref{Fig:3C-5} are indeed ordered and periodic, despite their Fourier-transformed trajectories of Fig. \ref{Fig:3C-6}(b) showing only weak frequency information. However, the initial SALI time evolution in Fig. \ref{Fig:3C-7aa}(a)-(b) for the pretension values of $F_{0}=\SI{30}{\newton}$ and $F_{0}=\SI{60}{\newton}$ has a sharp exponential decline hinting once again to transient chaotic behaviour.

\subsubsection{\text{Spring stiffness}}
\label{Sec:3C-4}

\begin{figure}[htbp]
\begin{centering}
\begin{overpic}[width=1.00\textwidth]{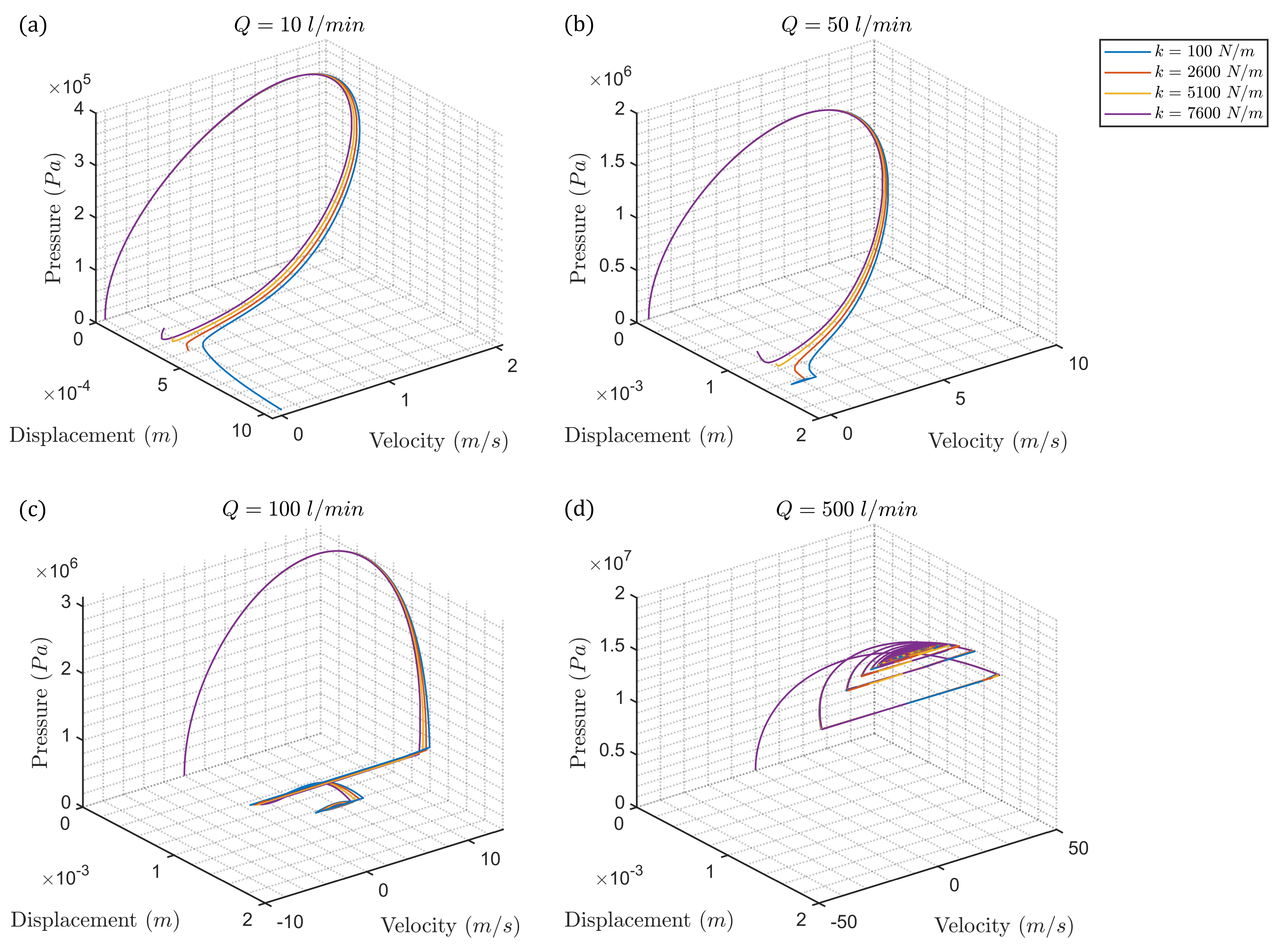}
\end{overpic}
\par
\end{centering}
\caption{Phase space with trajectories for the system's variables at various values of the spring stiffness and at flow rates of $Q=\SI{10}{\liter\per\minute}$ (a), $Q=\SI{50}{\liter\per\minute}$ (b), $Q=\SI{100}{\liter\per\minute}$ (c), and $Q=\SI{500}{\liter\per\minute}$ (d).}
\label{Fig:3C-8} 
\end{figure}

\begin{figure}[htbp]
\begin{centering}
\begin{overpic}[width=1.00\textwidth]{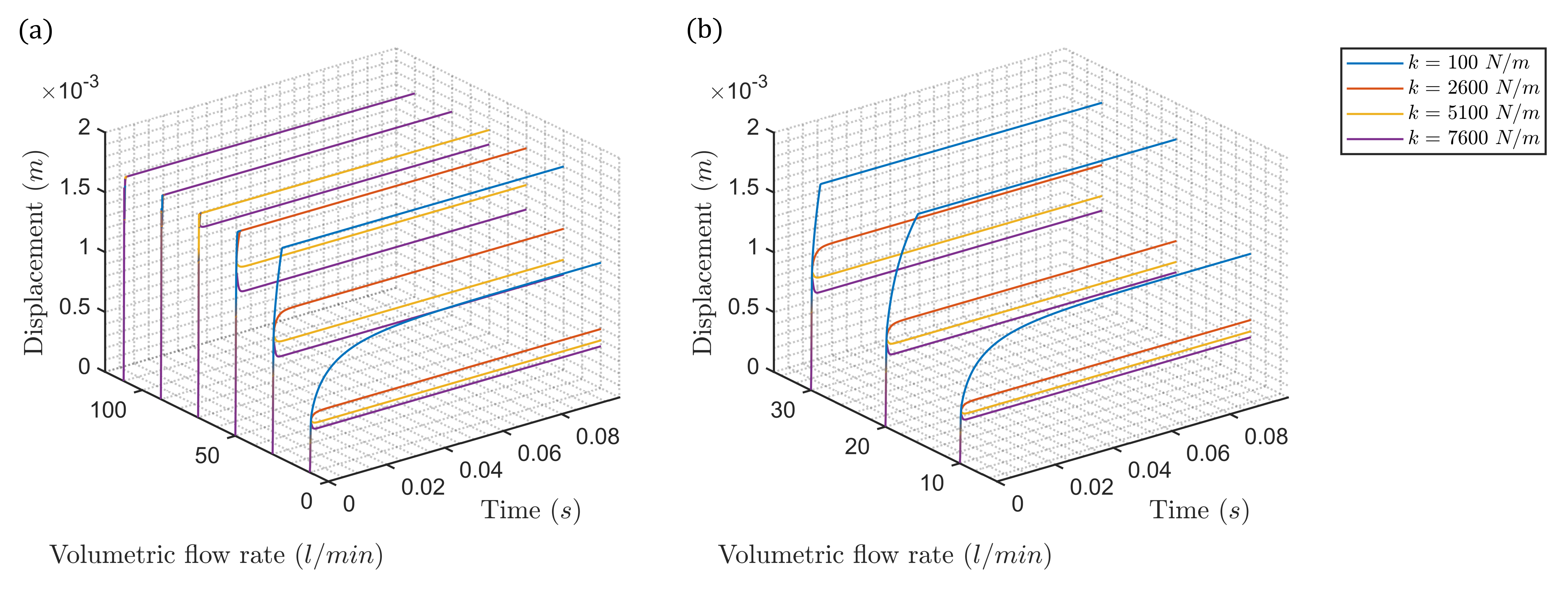}
\end{overpic}
\par
\end{centering}
\caption{Time series of displacement data for various stiffness values and flow rate ranges of $0$--$\SI{100}{\liter\per\min}$ (a) and $0$--$\SI{50}{\liter\per\min}$ (b).}
\label{Fig:3C-9} 
\end{figure}

\begin{figure}[htbp]
\begin{centering}
\begin{overpic}[width=1.00\textwidth]{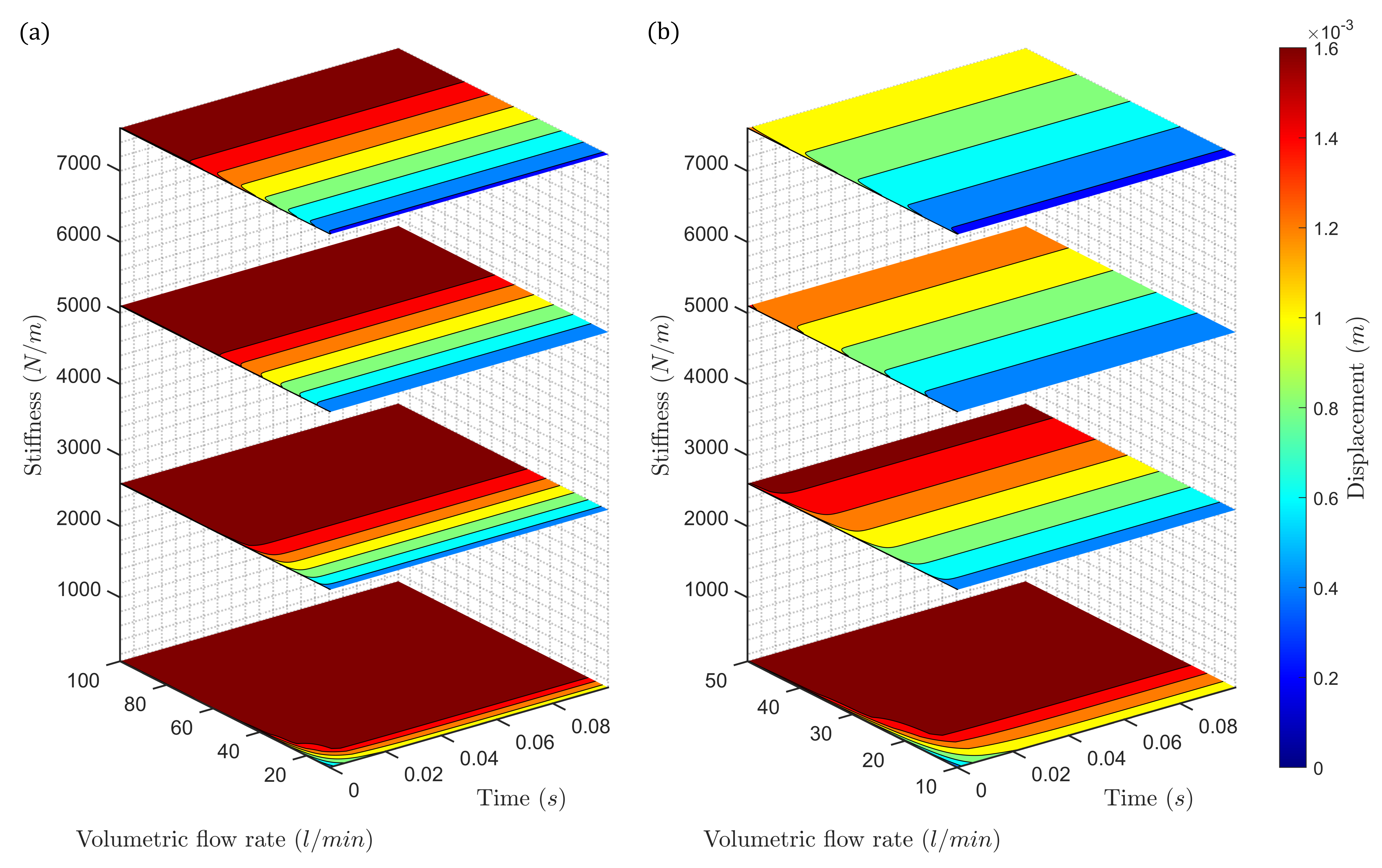}
\end{overpic}
\par
\end{centering}
\caption{Parameter space of displacement as a function of stiffness, time, and flow rate ranges of $0$--$\SI{100}{\liter\per\min}$ (a) and $0$--$\SI{50}{\liter\per\min}$ (b).}
\label{Fig:3C-10} 
\end{figure}

An increase in stiffness of the spring loading the valve shows a strong influence on the equilibrium solution to which the motion settles. However, the basic dynamics of the blowing-off or popping-off process is not affected (cf. Fig. \ref{Fig:3C-8}).
In particular, there is a clear initial overshooting of trajectories at intermediate to high stiffnesses of around $\SI{5100}{\newton\per\meter}$ to $\SI{7600}{\newton\per\meter}$, as visible in Fig. \ref{Fig:3C-9}. This is probably due to a delayed reaction of the valve to the stronger spring forcing.
Furthermore, Fig. \ref{Fig:3C-10} shows a gradual diminution of the opening amplitude at increasing values of spring stiffness, especially at small to intermediate flow rates up to approximately $\SI{70}{\liter\per\minute}$.

\subsubsection{\text{Damping}}
\label{Sec:3C-5}

\begin{figure}[htbp]
\begin{centering}
\begin{overpic}[width=1.00\textwidth]{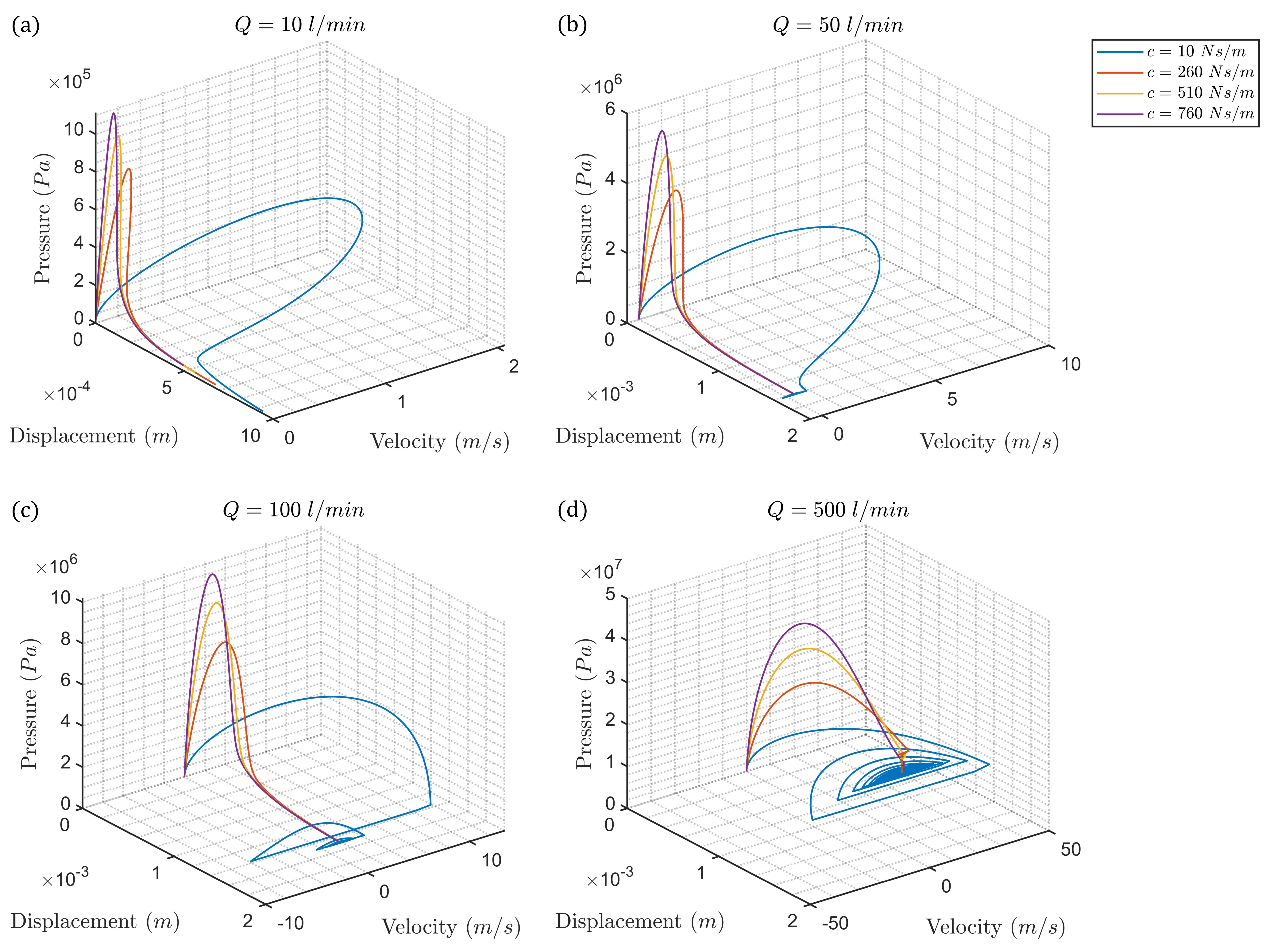}
\end{overpic}
\par
\end{centering}
\caption{Orbits in phase space for various damping coefficients at volumetric flow rates of $Q=\SI{10}{\liter\per\minute}$ (a), $Q=\SI{50}{\liter\per\minute}$ (b), $Q=\SI{100}{\liter\per\minute}$ (c), and $Q=\SI{500}{\liter\per\minute}$ (d).}
\label{Fig:3C-11} 
\end{figure}

\begin{figure}[htbp]
\begin{centering}
\begin{overpic}[width=1.00\textwidth]{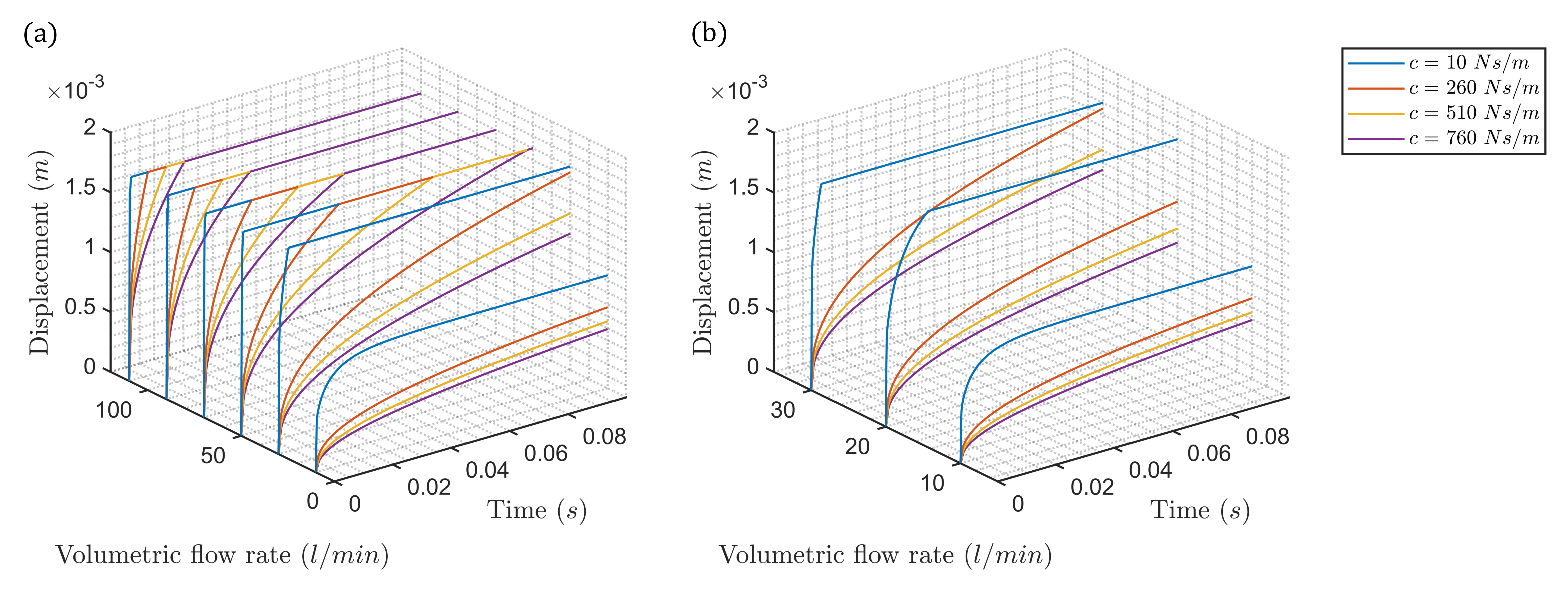}
\end{overpic}
\par
\end{centering}
\caption{Displacement time series for increasing values of the damping coefficient and flow rates of $0$--$\SI{100}{\liter\per\min}$ (a) and $0$--$\SI{50}{\liter\per\min}$ (b).}
\label{Fig:3C-12} 
\end{figure}

\begin{figure}[htbp]
\begin{centering}
\begin{overpic}[width=1.00\textwidth]{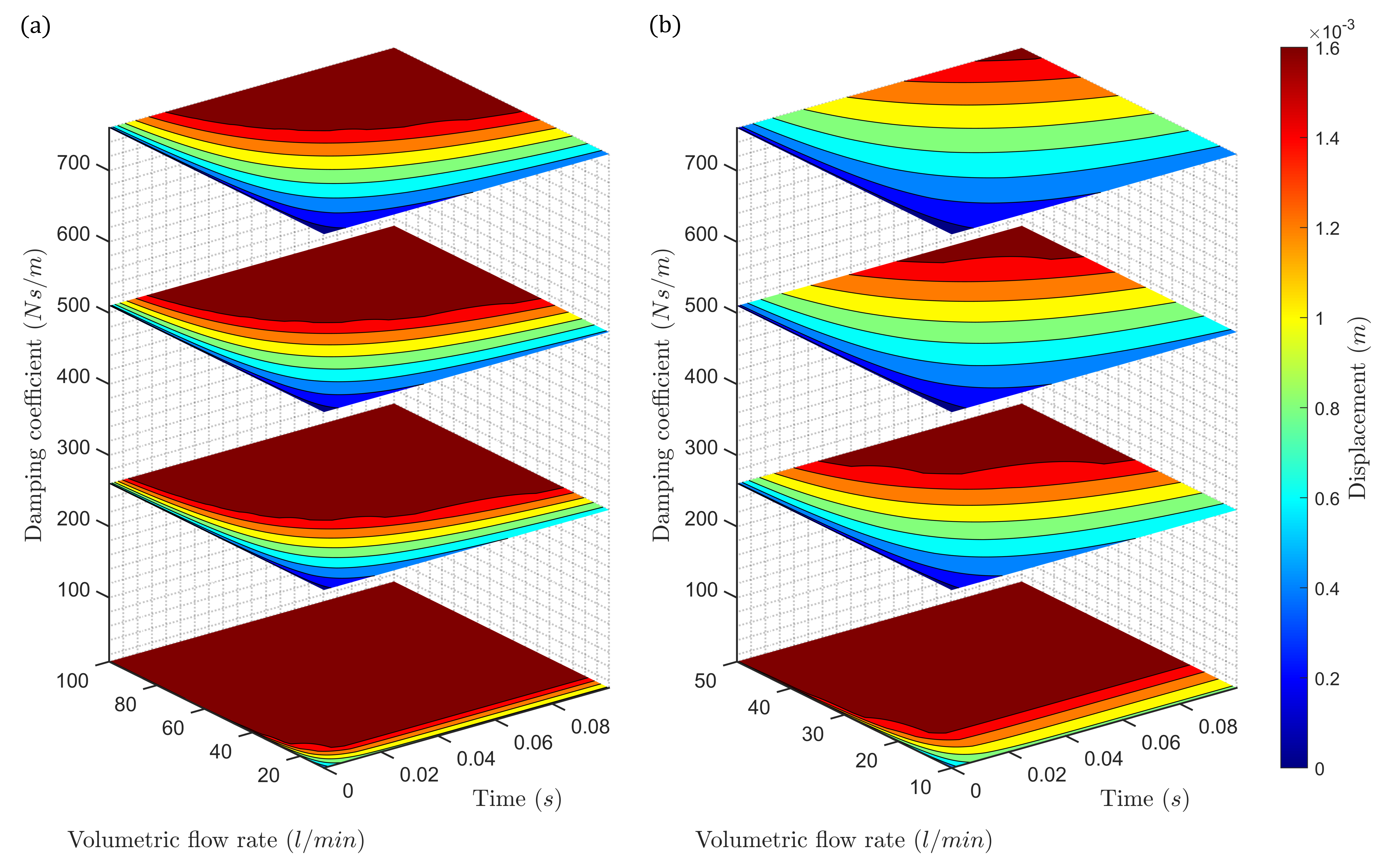}
\end{overpic}
\par
\end{centering}
\caption{Displacement in a parameter space of damping coefficient, time, and volumetric flow rate of $0$--$\SI{100}{\liter\per\min}$ (a) and $0$--$\SI{50}{\liter\per\min}$ (b).}
\label{Fig:3C-13} 
\end{figure}

Both structural and fluid damping is taken into account in the resulting damping coefficient of the system. This makes it difficult to quantify its exact value through experiments. In the following, the impact of its typical parameter ranges on the valve dynamics is investigated. Figure \ref{Fig:3C-11} shows a significant reduction of velocity already at intermediate values, pushing the orbits into displacement-pressure planes. Moreover, any chattering motion, which is significant at low damping of $c<\SI{100}{\newton\second\per\meter}$, effectively vanishes at $c>\SI{200}{\newton\second\per\meter}$, even for large volumetric flow rates. It can be recognized in Fig. \ref{Fig:3C-12} that the valve opening is also significantly delayed at higher damping, causing a shift in displacement amplitudes and opening times. This effect is occurring at all flow rates, but more severely at the lower range (cf. Fig. \ref{Fig:3C-13}).

\subsubsection{\text{Compressibility}}
\label{Sec:3D-6}

\begin{figure}[htbp]
\begin{centering}
\begin{overpic}[width=1.00\textwidth]{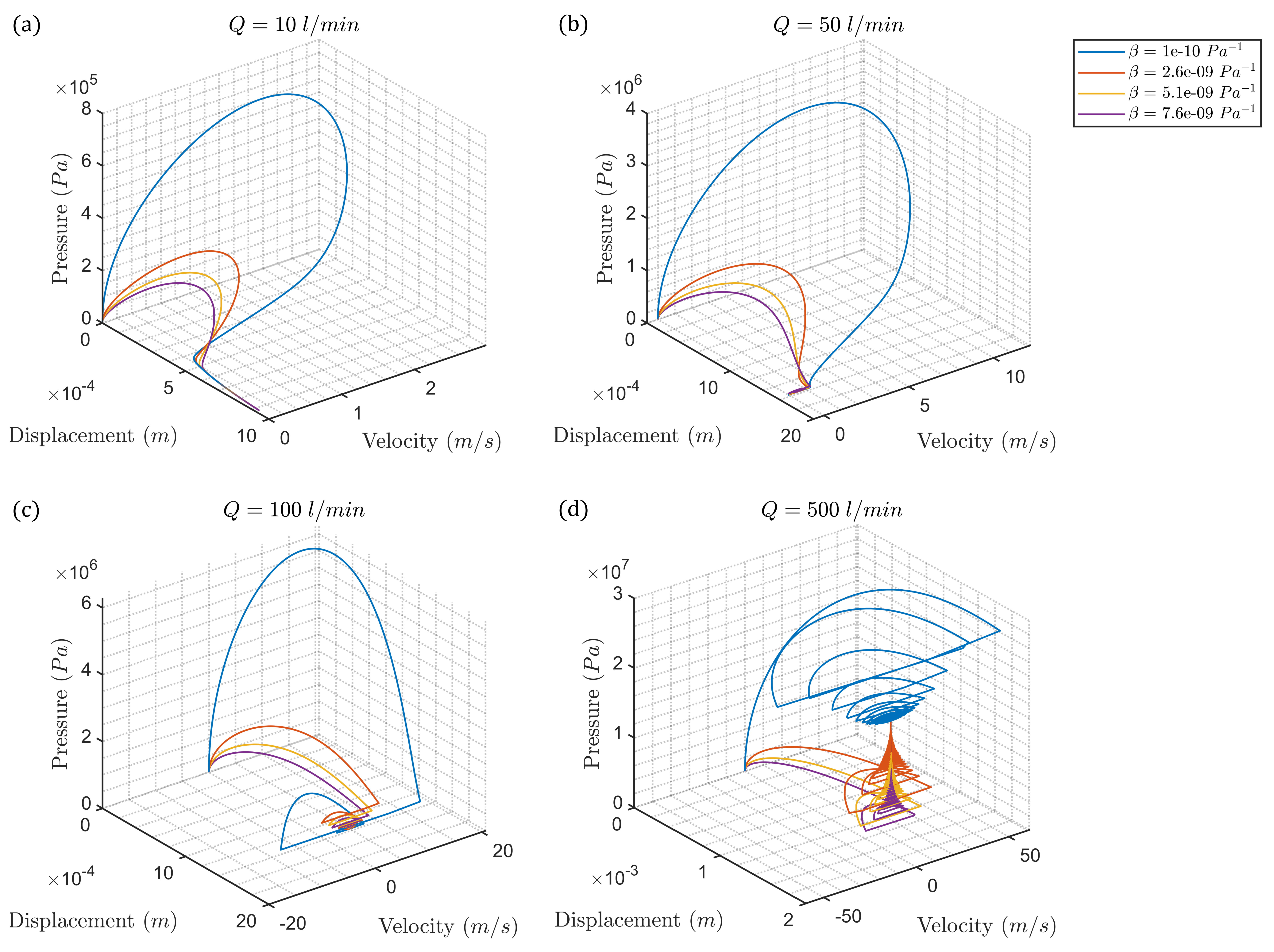}
\end{overpic}
\par
\end{centering}
\caption{Orbits in phase space for discrete values of compressibility at volumetric flow rates of $Q=\SI{10}{\liter\per\minute}$ (a), $Q=\SI{50}{\liter\per\minute}$ (b), $Q=\SI{100}{\liter\per\minute}$ (c), and $Q=\SI{500}{\liter\per\minute}$ (d).}
\label{Fig:3C-14} 
\end{figure}

A change in compressibility is the most likely reason for strong hysteresis effects of pressure relief valves in shock absorbers, as described in some detail by \citet{duym1997physical}.
The compressibility of hydraulic oils can increase due to several reasons such as oil quality, thermal expansion, or dissolved gas phase in the form of air bubbles. Particularly the presence of air in the mineral oil can alter the nominal resistance to compression. As Fig. \ref{Fig:3C-14} shows, higher compressibility clearly leads to a damping effect on all variables, introducing unpredictability and irreversibility to the system. This is one of the reasons why resulting damping force characteristics of the shock absorber do not coincide with compression and rebound strokes, despite comparable valves for both phases.

\section{\textbf{Conclusions}}
\label{Sec:4}

The basic function of the shim valves within a shock absorber resembles that of an impact oscillator, a non-smooth dynamical system that shows a multitude of nonlinear effects such as grazing bifurcations and chaotic motion \citep{di2008bifurcations}. 
Such effects can be detrimental and can compromise the regular function of the shock absorber by drastically increased response times or even complete loss of damping at high excitation frequencies and flow rates through the valve system. These effects are further complicated by the presence of multiple phases in the shock absorber (mineral oil in the main chamber, nitrogen in the gas reservoir). Due to this inherently nonlinear behaviour, it is of vital importance to understand the impact that parameter changes have on the dynamic features of a shock absorber pressure relief valve and on the nature of its motion (i.e. steady, periodic, aperiodic, or chaotic).

Regarding the main research questions outlined in Sec. \ref{Sec:1}, the following findings are documented:
\begin{itemize}
\item The quantitative value of shock absorber variables, such as valve displacement, velocity, and pressure, can be computed at its critical points for any set of boundary conditions and system properties by using the analytical expressions of Eq. (\ref{Eq:2A-15a}) and Eq. (\ref{Eq:2A-13}) for the case without impact ($F_{i}=0$) and Eq. (\ref{Eq:2A-18}) for the case with impact ($F_{i}\neq0$), where the fixed points are pseudo-equilibria defined by the bounds of the domain.
\item 
It is shown in Sec. \ref{Sec:3C} that while certain parameters, such as the bleed orifice cross-sectional area have a predictable, linear effect of the motion of the valve, other parameters, such as the mass and pretension of the spring-loaded valve show a tendency to destabilize the system under the current setup. 
This effect of the pretension force has also been noted by \citet{hHos2015dynamic} with respect to instability mechanisms of spring-loaded pressure relief valves in gas service. Although shock absorbers operate with a viscous fluid and are therefore highly dissipative, we could detect instability towards transient chaos in analogy to this study.
An increase in valve mass leads to a notable occurrence of low to medium frequency tonality in the range of $400$--$\SI{1200}{\hertz}$ and at medium to high volumetric flow rates of $40$--$\SI{500}{\hertz}$ (cf. Sec. \ref{Sec:3C-2}).
In contrast, the application of a pretension force leads to sustained periodic and non-periodic motion (cf. Sec. \ref{Sec:3C-3}).
\item The oscillatory behaviour summarized above is further analysed by using chaotic indicators, which quantify the level of order and chaos in the system and provide a measure of the time needed for transient chaos to subside. Moreover, Fourier transformations alone already give a good indication of the periodicity of the valve motion.
It is shown that intermediate values of pretension, as mentioned above, lead initially to aperiodic, transient chaotic valve dynamics, which are eventually damped out due to the high dissipation of the investigated system.
The chaotic indicator SALI is shown to be a potent method for the detection of transitory chaos in dissipative systems, such as the one studied here, due to its ability to gauge the immediate dynamical state, as opposed to the indication of  long-term behaviour by the Lyapunov exponents.
\end{itemize}

Based on the above conclusions, it becomes clear which quantities and parameter ranges lead to unsteady, vibratory valve response and to potentially dangerous loss of function due to instabilities. The results presented in Sec. \ref{Sec:3} are a direct extension to the study of the non-dimensional valve model by \citet{schickhofer2022fluid}, which investigated the exact onset of instabilities and Hopf bifurcations for realistic flow rates by application of the Routh-Hurwitz criterion.

Direct conclusions for the effective design of common shock absorber valves can be drawn from the data presented in this paper:
The valve mass should be kept as low as possible to ensure dynamic stability throughout the entire flow rate regime. Additionally, if pretension (e.g. by springs) is applied, this should be carefully adjusted for the operating flow rate, such that no strong valve oscillations are excited due to pretension and impact rebound force being of the same order of magnitude.
Consequently, by taking into account the global stability results from Tab. \ref{Tab:1} and Tab. \ref{Tab:2}, one can choose a balanced setup between mass, pretension, and target flow rate. Finally, the approach described in Sec. \ref{Sec:2} can be applied to other hydraulic valve systems to gain valuable information on their stability and nonlinear dynamics.

\bigskip

\noindent
\textbf{Acknowledgements} \\
This work has been performed under the Project HPC-EUROPA (INFRAIA-2016-1-730897), with the support of the European Commission for Research and Innovation under the H2020 Programme (Grant: HPC17ZLPYQ). In particular, the authors gratefully acknowledge the support of the Department of Mathematical Sciences at the University of Essex and the computer resources and technical support provided by the Irish Centre for High-End Computing (ICHEC).
\bigskip

\noindent
\textbf{Conflict of interest statement} \\
The authors have nothing to disclose that would have biased this work.

\newpage

\appendix

\section{Baseline system properties}
\label{App:A}

\setcounter{table}{0}
\setcounter{equation}{0}

Below the values for the model quantities of the shock absorber valve are given. They are based on a realistic pressure relief valve setup for the purpose of flow control in a suspension system. Wherever different values are chosen, this is explicitly stated in the text.

\begin{table}[htbp]
\caption{Properties of the valve system.}
\centering
\begin{tabular}{@{}lllll@{}}
\toprule
Parameter & Symbol & Value & Unit \\
\midrule
Mass & $m$ & $6.8768\times10^{-4}$ & $\SI{}{\kilogram}$ \\
Spring constant & $k$ & $200$ & $\SI{}{\newton\per\meter}$ \\
Impact stiffness & $k_{i}$ & $10^{10}$ & $\SI{}{\newton\per\meter}$ \\
Damping coefficient & $c$ & $10$ & $\SI{}{\newton\second\per\meter}$ \\
Pretension force & $F_{0}$ & $0.35$ & $\SI{}{\newton}$ \\
Pressure area & $A_{p}$ & $7.2150\times10^{-5}$ & $\SI{}{\square\meter}$ \\
Bleed orifice section & $A_{b}$ & $2.7\times10^{-6}$ & $\SI{}{\square\meter}$ \\
Discharge coefficient of valve opening& $C_{d,max,v}$ & $0.6$ & $-$ \\
Discharge coefficient of bleed orifice& $C_{d,max,b}$ & $0.6$ & $-$ \\
Momentum coefficient & $C_{f}$ & $0.3$ & $-$ \\
\bottomrule
\end{tabular}
\label{Tab:AA-1}
\end{table}

\begin{table}[htbp]
\caption{Properties of the mineral oil.}
\centering
\begin{tabular}{@{}lllll@{}}
\toprule
Parameter & Symbol & Value & Unit \\
\midrule
Reference density & $\rho_{0}$ & $830$ & $\SI{}{\kilogram\per\cubic\meter}$ \\
Reference compressibility & $\beta_{0}$ & $7.6\times10^{-10}$ & $\SI{}{\per\pascal}$ \\
Kinematic viscosity & $\nu$ & $2.3\times10^{-5}$ & $\SI{}{\square\meter\per\s}$ \\
Control volume& $V$ & $1\times10^{-4}$ & $\SI{}{\cubic\meter}$ \\
\bottomrule
\end{tabular}
\label{Tab:AA-2}
\end{table}

\newpage

\end{document}